\documentclass[11pt]{article}

\pdfoutput=1

\usepackage{a4wide,amssymb,cite,graphicx}
\usepackage{epsfig}
\usepackage[usenames,dvipsnames]{color}
\usepackage{slashed}
\usepackage{amsmath}

\usepackage{slashed}
\usepackage{amsmath,bm,bbm}
\usepackage{amsfonts}
\usepackage[titletoc,title]{appendix}
\usepackage[small]{caption}
\usepackage[margin=1in]{geometry}
\usepackage[multiple]{footmisc}
\usepackage{mathtools}
\usepackage[nottoc]{tocbibind}
\usepackage{xcolor}

\usepackage{hyperref}

\newcommand{\be}{\begin{equation}}
\newcommand{\ee}{\end{equation}}
\newcommand{\bea}{\begin{eqnarray}}
\newcommand{\eea}{\end{eqnarray}}

\def\circa#1{\,\raise.3ex\hbox{$#1$\kern-.75em\lower1ex\hbox{$\sim$}}\,}

\begin{document}

\begin{titlepage}

 \begin{flushright}
CERN-TH-2024-031, DESY-24-031
\end{flushright}

\begin{centering}
{\huge {\bf  Nonthermal Heavy Dark Matter \\
\vskip .4cm from a First-Order Phase Transition}} \\

\vspace{1.5cm}

{\bf Gian F. Giudice\,$^{1}$, Hyun Min Lee\,$^{2}$, Alex Pomarol\,$^{3,4}$, Bibhushan Shakya\,$^{5}$  }
%\\
\vspace{.5cm}

{\it $^1$CERN, Theory Department, 1211 Geneva 23, Switzerland. }
\\ \vspace{0.1cm}
{\it $^2$Department of Physics, Chung-Ang University, Seoul 06974, Korea.} 
\\ \vspace{0.1cm}
{\it $^3$IFAE and BIST, Universitat Aut`onoma de Barcelona, 08193 Bellaterra, Barcelona}
\\ \vspace{0.1cm}
{\it $^4$Departament de Fisica, Universitat Autonoma de Barcelona, 08193 Bellaterra, Barcelona}
\\ \vspace{0.1cm}
{\it $^5$Deutsches Elektronen-Synchrotron DESY, Notkestr.\,85, 22607 Hamburg, Germany}

\end{centering}
\vspace{1cm}

\begin{abstract}
We study nonthermal production of heavy dark matter from the dynamics of the background scalar field during a first-order phase transition, predominantly from bubble collisions. In scenarios where bubble walls achieve runaway behavior and get boosted to very high energies, we find that it is possible to produce dark matter with mass several orders of magnitude above the symmetry breaking scale or the highest temperature ever reached by the thermal plasma.  We also demonstrate that the existing formalism for calculating particle production from bubble dynamics in a first-order phase transition is not gauge invariant, and can lead to spurious results. While a rigorous and complete resolution of this problem is still lacking, we provide a practical prescription for the computation that avoids unphysical contributions and should provide reliable order-of-magnitude estimates of this effect. Furthermore, we point out the importance of three-body decays of the background field excitations into scalars and gauge bosons, which provide the dominant contributions at energy scales above the scale of symmetry breaking. Using our improved results, we find that scalar, fermion, and vector dark matter are all viable across a large range of mass scales, from $\mathcal{O}(10)$ TeV to a few orders of magnitude below the Planck scale, and the corresponding phase transitions can be probed with current and future gravitational wave experiments.
\end{abstract}

\vspace{3cm}

\end{titlepage}

\tableofcontents

\newpage

\section{Motivation}

The topic of first-order phase transitions (FOPTs) \cite{Hogan:1983ixn,Witten:1984rs,Hogan:1986qda,Kosowsky:1991ua,Kosowsky:1992rz,Kosowsky:1992vn,Kamionkowski:1993fg} -- where the metastable (false, unbroken) vacuum of the early Universe decays into its stable (true, broken) configuration through the nucleation, expansion, and percolation of bubbles of true vacuum -- has received intense scrutiny in the community in recent years. Studied extensively in the context of inflation \cite{Guth1981,LaSteinhardt1989} several decades ago, this phenomenon has become the subject of renewed interest due to its promise as a viable cosmological source of gravitational waves (GWs) \cite{Grojean:2006bp,Caprini:2015zlo,Caprini:2018mtu,Caprini:2019egz,Athron:2023xlk} that can be detected with current and upcoming gravitational wave experiments. Although the Standard Model (SM) of particle physics in its current form does not feature any FOPTs, such transitions can be readily realized  in many realistic beyond the Standard Model (BSM) scenarios \cite{Randall:2006py,Schwaller:2015tja,Jaeckel:2016jlh,Dev:2016feu,Baldes:2017rcu,Tsumura:2017knk,Okada:2018xdh,Croon:2018erz,Baldes:2018emh,Prokopec:2018tnq,Bai:2018dxf,Breitbach:2018ddu, Fairbairn:2019xog, Helmboldt:2019pan,Ertas:2021xeh,Jinno:2022fom} and therefore are of great interest. 

In this paper, we will be interested in FOPT processes that also address one of the most glaring shortcomings of the SM of particle physics: the identity of dark matter (DM). We will focus, in particular, on scenarios where the dynamics of the background field during the FOPT process is also responsible for DM production. Several papers in the literature have explored various qualitatively different realizations of this prospect. Refs.\!\cite{Baker:2019ndr,Chway:2019kft} examined configurations where a pre-existing thermal DM abundance in the unbroken phase can be filtered into the broken phase by slow-moving bubble walls to realize the correct exponentially suppressed relic abundance. Refs.\!\cite{Asadi:2021yml,Asadi:2021pwo,Asadi:2022vkc,Gehrman:2023qjn} made use of the trapping of dark sector particles in the false vacuum bubbles to realize the correct dark matter abundance. Ref.\!\cite{Azatov:2021ifm,Baldes:2022oev,Ai:2024ikj} studied cases where particles crossing across relativistic bubble walls can upscatter into heavy states that are, or can produce, DM. Ref.\!\cite{Hambye:2018qjv,Baratella:2018pxi,Baldes:2021aph} studied frameworks where the dynamcis associated with supercooled transitions in a confining sector produce the correct DM abundance. Ref.\!\cite{Falkowski:2012fb} explored the prospects of producing heavy DM from bubble collisions in an electroweak phase transition, finding that scalar DM cannot be produced with the desired abundance but the production of heavy vector or fermion DM is possible; Ref.\!\cite{Freese:2023fcr} extended this idea to DM in a dark sector. Ref.\!\cite{Baldes:2023fsp} considered DM production from the collisions of shells of boosted particles around the bubble walls. 

Note that all of the above ideas (except \cite{Falkowski:2012fb,Freese:2023fcr}) rely on interactions between bubble walls and particles in the ambient thermal bath. In this paper, we focus on DM production from the spacetime dynamics of the background scalar field itself as it undergoes various stages of the phase transition. This is a fundamental, unavoidable contribution that is present in \textit{any} FOPT (including all of the above cases), irrespective of the existence or nature of a thermal bath of particles. Particle production from a changing background field is a well-known physical phenomenon familiar from various contexts, such as gravitational particle production \cite{Parker:1969au,Grib:1969ruc,Zeldovich:1971mw,Kolb:2023ydq} (for some specific applications to dark matter production, see e.g.~\cite{Aoki:2022dzd,Lee:2023dcy}), Schwinger effect \cite{Schwinger:1951nm}, and Hawking radiation from black holes \cite{Hawking:1974rv,Hawking:1975vcx}. The calculation of particle production from background field dynamics during various stages of a FOPT, in particular from bubble collisions, is complicated due to the inhomogeneous nature of the process, but can be calculated in a manner analogous to the production of gravitational waves. The formalism to study this process was first developed in \cite{Watkins:1991zt} in the context of reheating after  first-order inflation. This formalism was then explored by \cite{Konstandin:2011ds} for cold baryogenesis, and further developed in \cite{Falkowski:2012fb} with semi-analytic results for some idealized bubble collision cases and applications for nonthermal DM production. More recently, the results were refined with numerical studies for more realistic bubble collisions in \cite{Mansour:2023fwj}, and various aspects of the underlying physics clarified in \cite{Shakya:2023kjf}. Particle production from bubble collisions was shown to be a new source of GWs in \cite{Inomata:2024rkt}. In this paper, we use the improved results from \cite{Mansour:2023fwj,Shakya:2023kjf} to calculate the production of DM from the background field dynamics in a dark phase transition. 

While the DM production mechanism we discuss here is very general, it is particularly well-suited for the production of heavy DM with mass far above the scale of symmetry breaking or the temperature of the plasma following the phase transition. Ultraheavy DM, with mass ranging from $\mathcal{O}(100)$ TeV to the Planck scale, is of broad interest to the community for several theoretical as well as experimental reasons \cite{Carney:2022gse}, but generally suffers from the lack of viable production mechanisms that realize the correct relic density. Recall that if the Universe reheats to temperatures comparable to the DM mass (in particular, above the freezeout temperature of DM), DM will rethermalize with the bath, erasing the effects of earlier cosmological events (such as bubble dynamics), and will undergo conventional freezeout, which cannot produce the correct relic abundance for DM masses above the unitarity bound of $\mathcal {O}(100)$ TeV. Hence nonthermal production mechanisms are required for the production of ultraheavy DM. Nonthermal freeze-in requires extremely small ($\sim10^{-10}$) couplings \cite{McDonald:2001vt,Hall:2009bx}, whereas production at temperatures lower than the DM mass suffers from exponential (Boltzmann) suppression. In this context, FOPTs provide a unique configuration not found in other cosmological setups that can nonthermally produce extremely heavy particles with masses significantly larger than any other energy scale achieved in the early Universe: since the bubble walls can accelerate to relativistic speeds in the absence of friction from the plasma, they can reach energies far above the energy scale of the phase transition or the temperature of the ambient plasma. The possibility of producing particles with masses far greater than the scale of phase transition from the collisions of such bubbles was recognized in \cite{Watkins:1991zt}, and subsequently employed for heavy DM production in \cite{Falkowski:2012fb,Freese:2023fcr}; we will extend these studies, clarifying and improving on several important aspects.  We will discuss scalar, fermion, and vector DM, highlighting qualitatively distinct features and novel developments relative to the existing literature in each case. Furthermore, since FOPTs from dark sectors can give rise to large gravitational wave signals, such configurations provide an opportunity to detect this production mechanism for DM in dark sectors (otherwise inaccessible with other experimental probes) with gravitational waves, providing added motivation for this study.

This paper also contains two important developments on the formalism to calculate particle production. First, we demonstrate that the existing formalism is gauge-dependent and leads to spurious results in certain cases if not treated carefully. While we are unable to provide a complete resolution of this problem, we provide a practical prescription for the computation that extracts physical contributions and can be reliably used for order-of-magnitude estimates of this effect. Second, we point out that for the production of gauge bosons and scalars, three-body decays of the background field excitations (rather than two-body decays, which is the only contribution currently considered in the literature) provide the dominant contribution for heavy DM. Both results are crucial and substantially change the conclusions regarding the viability and parameter space for DM derived in previous works. 

This paper is organized as follows. In Sec.\,\ref{sec:framework}, we describe the framework for the study, discussing the relevant phase transition configurations, parameters, and particle content.
The formalism for the calculation of particle production from various stages of a FOPT is described in Sec.\,\ref{sec:formalism}. Sec.\,\ref{sec:gaugedependence} discusses issues related to the gauge dependence of the formalism, and provides a practical solution to the problem that enables the calculation of scalar and gauge boson production, including the three-body configurations that provide the dominant contributions. In Sec.\,\ref{sec:dmproduction}, we discuss dark matter production contributions from other processes before, during, and after the phase transition, as well as subsequent evolution of the DM population. The parameter space where DM can be produced with the correct relic abundance from FOPTs is presented in Sec.\,\ref{sec:parameterspace}. Sec.\,\ref{sec:summary} contains a summary of the main results of this paper and a discussion of various related ideas. 

\section{Framework}
\label{sec:framework}
The dark matter production from FOPT background field dynamics that we calculate in this paper is very generic: it occurs unavoidably at any FOPT if the DM particles couples directly or indirectly to the background field undergoing the phase transition, independently of the details of the FOPT or the thermal plasma. We therefore present our analysis and results in a ``model independent" manner, in terms of phenomenologically relevant parameters characterizing the phase transition and for simplified minimal DM setups, so that the results can be applied in a straightforward manner to specific dark sector and DM models. 

\subsection{Phase Transition Parameters}
\label{parameters}

Here, we list the phenomenological parameters that are relevant for the calculation. Consider a FOPT in a dark/hidden sector where a background field $\phi$ transitions from a metastable, false vacuum, where it has a vanishing vacuum expectation value (vev) $\langle \phi \rangle=0$, to a stable, true vacuum configuration with non-vanishing vev $\langle \phi \rangle=v_\phi$. The latent energy released in the phase transition is given by the difference in the potential energies of the two vacua, and we parameterize it as 
\be
\Delta V \equiv V_{\langle \phi \rangle=0}-V_{\langle \phi \rangle=v_\phi}=c_V\,v_\phi^4\,.
\ee

The phase transition  parameters relevant to our calculation of DM production and abundance are:

\begin{itemize}

\item $T_n$:  temperature of the thermal bath at which the FOPT is triggered, i.e.\,when bubbles of true vacuum begin to nucleate at a rate greater than the Hubble scale. 

\item $R_0$: critical radius of nucleated bubble that can grow. This is typically $\mathcal{O}(T_n^{-1})$. 

\item $\alpha$:  strength of the phase transition, defined as $\alpha\equiv\frac{\rho(\text{vacuum})}{\rho(\text{radiation})}$, where $\rho(\text{vacuum})=\Delta V$ and $\rho(\text{radiation})$ represents the energy density in the radiation bath (SM and dark sectors combined) at $T_n$. 

\item $\beta$:  (inverse) duration of phase transition. This is generally parametrized relative to the Hubble scale as $\beta/H$, which is a dimensionless parameter. 

\item $v_w$: velocity of the bubble wall. This quantity is time-dependent: as the bubbles expand, vacuum energy gets transferred to the wall, accelerating it. Hence $v_w$ tends to grow, but can asymptote to a constant value in the presence of significant frictional forces. 

\item $\gamma_w$: Lorentz boost factor of the bubble wall, determined from $v_w$ via the relation $\gamma_w=1/\sqrt{1-v_w^2}$. In this paper we are interested in the relativistic regime $v_w\approx 1,~\gamma_w\gg 1$. 

\item $l_w$: thickness of the bubble wall. This quantity is also time dependent: while the wall thickness at bubble nucleation is $l_{w_0}\sim\mathcal{O}(v_\phi^{-1})$, the $\textit{apparent}$ wall thickness in the plasma frame gets Lorentz contracted as the bubble accelerates to greater velocities, hence $l_w=l_{w_0}/\gamma_w$ tends to decrease with time.

\item $R_*$:  typical size of vacuum bubbles at collision; this is determined from the timescale over which the transition completes, $R_*\approx v_w\,(8\pi)^{1/3}{\beta}^{-1}$.

\item $T_*$:  temperature of the thermal bath at which bubbles of true vacuum percolate and the phase transition ends. Since phase transitions complete within a fraction of Hubble time, $T_*\approx T_n$ if the Universe remains radiation dominated throughout. If the Universe instead becomes vacuum dominated, then $T_*$ is determined through energy conservation conditions at the end of the transition. 

\end{itemize}

In a specific model, these quantities can be calculated from the parameters in the underlying theory, as described in detail in several extensive reviews of phase transitions (see e.g.~\cite{Grojean:2006bp,Caprini:2015zlo,Caprini:2018mtu,Caprini:2019egz,Athron:2023xlk}). For our purposes, we will treat them as independent parameters (except for the relations described above), so that it should be straightforward to map our results to any given model by calculating the corresponding parameters in the model. 

\subsection{(Runaway) Phase Transition Configurations}
\label{subsec:runaway}

We now discuss the phase transition setups that are relevant for ultraheavy DM production. As stated earlier, we are particularly interested in scenarios where the DM mass is higher than the scale of the phase transition as well as the temperature of the thermal bath. This requires the bubble walls to gain sufficient energy to produce the heavy DM particles. We are thus interested in configurations where the bubble walls achieve so-called ``runaway" behavior, i.e.~are not slowed down by friction effects but continue to accelerate as they gain the latent energy in the false vacuum released in the transition. As we discuss here, whether this occurs depends on the details of the contents of the thermal bath as well as the nature of the transition. 

If the dark sector is in thermal equilibrium with the SM bath at some point in the early Universe, the two sectors share the same temperature, which remains true after the two sectors decouple (up to small corrections). However, this is not necessary, and the dark sector may be cold, i.e.~has a temperature substantially smaller than that of the SM bath (this occurs, for instance, if the inflaton or the lightest moduli fields preferentially reheat the visible (SM) sector), or hotter (in the opposite scenario). Requiring the two sectors to remain decoupled in this manner enforces an upper limit on possible portal couplings between the two sectors. This condition can be approximately quantified as $\lambda_{\text{portal}} \lesssim  \sqrt{T/M_{Pl}}$ for any temperature $T$ higher than the mass of the corresponding dark sector particle. Here $M_{Pl}$ is the Planck mass, and the portal coupling could be e.g.~a quartic coupling between the scalar $\phi$ and the SM Higgs boson, or a kinetic mixing between a dark gauge boson and the SM hypercharge. 

In general, a FOPT can occur either due to thermal effects (temperature dependent corrections to the scalar potential causes the true vacuum to become energetically favoured) or via quantum tunneling (the age of the Universe approaches the lifetime for the scalar field to tunnel into the true vacuum even with the zero temperature potential) -- see e.g.\!\cite{Fairbairn:2019xog} for detailed discussions. The former requires a thermal bath of hidden sector particles (which may or may not be in equilibrium with the SM bath), and a large coupling between the scalar field and some other particle in the bath; in this case, with $\mathcal{O}(1)$ couplings, the phase transition generally occurs at a temperature $T_*\sim v_\phi$ (if the dark and visible sectors have different temperatures, then it is the dark sector temperature that is relevant here), and is completed within a small fraction of Hubble time, $\beta/ H_*\sim 100-1000$ (where $H_*$ is the Hubble scale at temperature $T_*$ when the phase transition completes), as the bubble nucleation rate becomes extremely rapid once thermal corrections make the true vacuum energetically favorable. Transitions via quantum tunneling, on the other hand, do not require a thermal bath of hidden sector particles (i.e.~the hidden sector can be extremely cold before the transition, and the hidden sector energy density exists primarily in the form of vacuum energy), although it might be present;  the time of transition is determined by the shape of the potential. In such instances, $T_*$ can be several orders of magnitude smaller than $v_\phi$.  
Such transitions tend to last longer, completing in an $\mathcal{O}(1)$ fraction of Hubble time, so that $\beta/ H_*\sim 10$.

In either case, bubbles of true vacuum nucleate with critical radii $R_0$ and expand, accelerating as the latent energy released from the false vacuum is converted to kinetic and gradient energies in the bubble walls. Expanding bubbles encounter friction due to particles in the thermal bath crossing the wall and becoming massive in the broken phase. A full thermal distribution of a particle species crossing into the bubble is known to produces a pressure \cite{Bodeker:2009qy} (see also \cite{Dorsch:2018pat,Espinosa:2010hh,Ai:2024shx}):
\be
\mathcal{P}_{\text{LO}}\approx \frac{1}{24}m^2 T^2 \,,
\label{eq:pressure}
\ee
where $m$ is the mass of the particle in the broken phase and $T$ is the temperature of the bath. If the sum of such effects from all particles exceeds the energy available from the transition, $\Delta V$, the walls achieve a terminal velocity corresponding to some steady state configuration \footnote{For a thick-walled bubble,  $\Delta V$ should represent the difference in potential energies between the false vacuum and the field value to which the scalar field tunnels, rather than the true vacuum.}; if not, the walls continue to accelerate. As the walls become relativistic, friction due to splitting or transition radiation, corresponding to radiation of gauge bosons from particles crossing into the bubbles, becomes increasingly important \cite{Bodeker:2017cim,Hoche:2020ysm,Gouttenoire:2021kjv}, producing pressure that scales as
\be
\mathcal{P}_{\text{NLO}}\sim g^2\, \gamma_w\, m_V\, T^3 \,,
\label{eq:pressure2}
\ee
where $g$ is the gauge coupling and $m_V$ is now the mass of the gauge boson, and we have dropped some $\mathcal{O}(1)$ factors. This implies that the bubble walls reach a terminal velocity corresponding to $\gamma_w\sim \Delta V/(g^3 T^3 v_\phi)$ (where we have used $m_V=g v_\phi$) if they have not collided with other bubbles before this value is reached. 

If the frictional energy loss remains subdominant to $\Delta V$, energy conservation dictates that the boost factor of the wall grows with the growing bubble radius $R$ as $\gamma\approx \frac{2 R}{3 R_0}$ \cite{Ellis:2020nnr}. In such configurations, the boost factor can reach extremely large values; parametrically,
\be
\gamma_{\text{max}}\sim \frac{1}{\beta/H}\frac{M_{Pl}}{v_\phi}\,,
\label{eq:gammafactor}
\ee
where we have used the relations in Sec.\,\ref{parameters} and assumed $T\sim v_\phi$. The energy density in the bubble wall at collision is then $E_{\text{wall}}=\gamma_{\text{max}}/l_{w0}\sim M_{Pl}/(\beta/H)$, making it possible to produce heavy particles up to this scale. Remarkably, note that $E_{\text{wall}}$ is independent of $v_\phi$: a transition at a lower scale $v_\phi$, where the bubble walls have lower energy, is compensated  by a lower Hubble scale, which allows the bubbles to expand for longer before collisions occur, and thus the bubble walls can get boosted for a longer period. 

Viable scenarios that can realize such $\gamma\gg1$ runaway behavior needed for producing ultraheavy DM can broadly be classified into four distinct categories:

\vskip 0.1cm
\noindent\textit{Scenario I: Thermal transition without a gauge boson}

This corresponds to scenarios where the FOPT is thermally triggered, i.e.~a thermal bath that interacts with the bubble walls is present, but $\Delta V>\mathcal{P}_{\text{LO}}$, so that the friction from particles crossing into the bubbles and becoming massive is not sufficient to slow the walls down, and in the absence of a gauge boson there is no $\mathcal{P}_{\text{NLO}}$ contribution.

\vskip 0.1cm
\noindent\textit{Scenario II: Thermal transition with a light gauge boson}

Even if the broken symmetry is gauged, runaway behavior can be realized if the corresponding gauge boson is light $(m_V\ll v_\phi)$, i.e.~the gauge coupling is small ($g\ll1$). Recall that friction due to splitting radiation (Eq.\,\ref{eq:pressure2}), which grows linearly with $\gamma_w$, eventually saturates the released latent energy, resulting in a terminal value  $\gamma_w\sim \Delta V/(g^3 T^3 v_\phi)$ for the wall boost factor. Assuming $T\sim v_\phi$, we have $\gamma_w\sim c_V/g^3$, hence $\gamma_w\gg1$ is possible if $g\ll1$. In such cases, the boost factor at collision is 
\be
\gamma_w\sim \text{min}\left[\frac{c_V}{g^3},~\frac{2 R_*}{3 R_0}\right]\,,
\label{gammalightg}
\ee
i.e. either the terminal behavior described above is reached, or the bubble walls collide before this occurs.

\vskip 0.1cm
\noindent\textit{Scenario III: Supercooled phase transition}

Alternately, one could have a supercooled FOPT \cite{Randall:2006py,Konstandin:2011dr,vonHarling:2017yew,Ellis:2018mja,Baratella:2018pxi,Bruggisser:2018mrt,DelleRose:2019pgi,VonHarling:2019rgb,Fujikura:2019oyi,Ellis:2019oqb,Brdar:2019qut,Baldes:2020kam,Bruggisser:2022rdm}. In such transitions, $\Delta V > \rho_\text{radiation}$, leading to a period of vacuum domination and inflation that causes significant dilution of the pre-existing thermal bath before the phase transition completes.  The bubble walls therefore effectively expand in vacuum, encountering negligible friction, and can reach runaway behavior. In this case, note that reheating after the completion of the phase transition creates a thermal bath with $\rho_\text{radiation}\approx\Delta V$. 

\vskip 0.1cm
\noindent\textit{Scenario IV: Quantum tunneling in a cold dark sector}

Even if a dark thermal bath is effectively absent, the transition could occur via quantum tunneling; in this case, there are essentially no particles that interact with the bubble walls, and the walls continue to accelerate as the bubbles expand. The SM bath could be present or absent; if it is absent or its energy density is lower than the latent energy $\Delta V$ in the false vacuum, this leads to a vacuum dominated epoch, corresponding to the supercooled regime discussed above. For transitions that occur via quantum tunneling, the bubbles generally cannot percolate in a vacuum dominated inflating regime (this is essentially the graceful exit problem in first-order inflation models); to avoid this, we can assume for simplicity that the SM bath is present with energy density equal to or greater than the latent energy in the false vacuum, so that the Universe remains radiation dominated throughout and does not enter an inflationary phase during the phase transition. To draw the distinction with Scenario III above, by quantum tunneling we will therefore mean a transition that occurs in the absence of a dark sector bath, but without a supercooled (i.e.~vacuum-dominated) phase due to the dominance of the SM bath. 

\subsection{Particle Content }
\label{subsec:content}

As stated earlier, we will perform our analysis and calculations for DM production in simplified frameworks. We will assume the FOPT is characterized by a complex scalar field $\phi$ with vevs $\langle \phi\rangle =0,v_\phi$ and masses $m_\phi=m_f,m_t$ in the false (unbroken) and true (broken) vacua, with a self-interaction term $\frac{\lambda_\phi}{4!}|\phi|^4$. To minimize notation, we will also use $\phi$ to denote the physical component of the field (the radial mode, or the Higgs boson) later in the paper. The broken symmetry might be global or local; this implies the existence of either a massless Goldstone or massive gauge boson, respectively, in the broken theory. In the latter case, the mass of the gauge boson $Z'$ is  $m_{Z'}=g\,v_\phi$, where $g$ is the dark gauge coupling. We will assume that all of these dark sector particles (with the exception of the DM particle) can decay into the SM through small portal couplings, so that the energy in the dark sector eventually gets transferred to the SM bath. Such decays could be necessary to avoid overclosing the Universe or producing dark radiation that leads to excessively large contributions to the effective number of relativistic degrees of freedom $N_{\text{eff}}$ in the late Universe if the energy density in the dark sector is substantial. Such decays into the SM might not exist for the Goldstone, which can obtain a small mass due to quantum-gravity effects but might not have any SM decay channels kinematically accessible. In this case, one must ensure that the Goldstone accounts for less than roughly one percent of the total energy density in the Universe for consistency with $N_{\text{eff}}$.

For the DM particle $\chi$, we will examine scalar, fermion, as well as vector candidates, which we denote as $ \chi_s,~\chi_f,$ and $\chi_v$, respectively. If the DM mass is at or below the scale of symmetry breaking, $m_\chi\lesssim v_\phi$, DM could have obtained its mass during the FOPT from the $\phi$ vev. For ultraheavy masses $m_\chi\gg v_\phi$, which is the primary regime of interest to us, DM mass is generated at some heavy scale, and the dynamics of symmetry breaking associated with the FOPT has negligible effect on the DM mass. 

We will consider the following simplified interactions between the DM candidates and the background field:

\begin{itemize}
\item \textbf{Scalar DM  $\chi_s$}, with mass $m_{\chi_s}$ and interaction $\frac{\lambda_s}{4} |\phi|^2 \chi_s^2$. 

Note that this is a renormalizable operator that can be valid to arbitrarily high scales. Since the above interaction term produces a mass contribution $\sqrt{\lambda_s/2}\, v_\phi$ once $\phi$ obtains a nonzero vev, we will focus on the regime $m_{\chi_s}^2>\frac{1}{2}\lambda_s v_\phi^2$, and treat $m_{\chi_s}$ and  $\lambda_s$ as independent quantities for simplicity. 

\item \textbf{Fermion DM  $\chi_f$}, with mass $m_{\chi_f}$ and effective interaction $y_f \phi \chi_f\bar{\chi}_f$ (+h.c.). 

Here the $\phi \chi_f\bar{\chi}_f$ interaction implies that the $\chi_f\bar{\chi}_f$ combination is charged under the symmetry that is broken by the $\phi$ vev. Here $\chi_f$ could be a chiral fermion, obtaining its mass from the $\phi$ vev after the symmetry is broken, analogous to the fermions interacting with the Higgs field in the SM; however, in this case $m_{\chi_f}=y_f v_\phi \lesssim v_\phi$. Alternately, the effective $y_f \phi \chi_f\bar{\chi}_f$ interaction could have been derived from a higher dimensional operator of the form $\frac{y'_f}{\Lambda_f}|\phi|^2 \chi_f\bar{\chi}_f$, where $\Lambda_f$ is some ultraviolet (UV)-cutoff scale. In this case, $\chi_f$ does not have to carry any charge associated with $\phi$, and its mass can be significantly larger than the symmetry breaking scale of interest, $m_{\chi_f}\gg v_\phi$, and the effective coupling is $y_f=2 y'_f\, v_\phi/\Lambda_f$. A specific realization of this (see \cite{Falkowski:2012fb}) involves $\phi$ mixing with some singlet scalar $S$ that couples to the fermion $\chi_f$. Here we remain agnostic about such underlying details and simply work with the effective interaction term $y_f \phi \chi_f\bar{\chi}_f$. As with the scalar case, we will focus on masses larger than that obtained from the symmetry breaking, $m_{\chi_f}>y_f\,v_\phi$,  and consider $m_{\chi_f}$ and  $y_f$ as independent parameters. 

\item \textbf{Vector DM $\chi_v$}, with mass  $m_{\chi_v}$ and an interaction of the form $\frac{1}{2}\lambda_V |\phi|^2 (\chi_v)^\mu (\chi_{v})_{\mu}$.

Note that $(\chi_{v})_{\mu}$ is not the field strength tensor, but a component of the gauge field $\chi_v$, where the subscript $v$ is not a Lorentz index but simply denotes that this is a vector DM candidate (analogous to the notation in the previous bullet points, $\chi_s$ for scalar DM and $\chi_f$ for fermion DM). Thus the above interaction term  is a dimension 4 coupling between two scalar and two vector fields.
Again, this interaction does not necessitate that the gauge boson $\chi_v$ corresponds to the gauge symmetry broken by $\phi$, as it could arise from integrating out intermediate particles (e.g.~a singlet mediator field, see \cite{Falkowski:2012fb} for more detailed discussions). For a vector boson, additional subtleties arise from the interplay between its transverse and longitudinal modes; these aspects will be discussed in Sec.\,\ref{sec:vector}. As in the previous two cases, we will treat the mass and coupling as independent quantities.  
\end{itemize}

In all scenarios, we will restrict ourselves to cases where DM is heavier than the scalar and the gauge/Goldstone boson, i.e.~$m_\chi> m_f,m_t, m_{Z'},m_G$, so that DM cannot be produced from decays of other particles in the dark sector, otherwise it can be produced from the oscillations of the scalar field long after the bubble collisions, effectively reaching a thermal abundance, in which case it either re-establishes thermal equilibrium with the bath or tends to be overproduced and overclose the Universe. 

Note that the coupling of the scalar field $\phi$ to particles far heavier than its mass can produce radiative contributions that can lift its mass to the heavy scale, hence the hierarchy $m_\phi, v_\phi \ll m_\chi$ could involve significant fine-tuning. Likewise, the coupling between $\phi$ and DM can produce corrections to the scalar potential that could modify the nature of the phase transition; this is particularly concerning in scenarios where the coupling is large ($\mathcal{O}(1)$), or where the realization of the FOPT requires some amount of tuning. However, as we will see below, heavy dark matter production from bubble collisions can also be realized with extremely small couplings between $\phi$ and DM, as small as $\mathcal{O}(10^{-10})$, which do not alter the scalar potential appreciably enough to affect the FOPT. In any case, such concerns are best addressed in complete particle physics models, and we ignore such considerations in our simplified framework treatment in this paper. 

Finally, additional dark sector particles beyond the ones discussed above might exist, but their existence is irrelevant as long as they do not couple more strongly to DM than the scalar $\phi$ and do not produce significant effects on bubble wall dynamics; we will assume this to be the case for the purposes of this paper.

%%%%%%%%%%%%%%%%%%%%%%%%%%%

\section{Formalism: Particle Production Calculation}
\label{sec:formalism}

In this section, we describe the formalism for calculating particle production from the dynamics of the background field during a FOPT. The transition consists of three stages: bubble nucleation, expansion, and collision, all of which contribute to particle production, see \cite{Shakya:2023kjf} for detailed discussions. Although the contributions from the former two stages are subdominant for the production of heavy particles, we will discuss them here briefly for completion. 

\subsection{Bubble Nucleation}

In the thin-wall limit (where the thickness of bubble walls separating the true and false vacua is significantly smaller than the size of the nucleated bubble, $l_{w0}\ll R_0$), the dynamics of the background field within the bubble can be assumed to be homogeneous, and the number density of a particle species $Y$ produced within the bubble during the nucleation process can be estimated as \cite{Shakya:2023kjf}
\be
n_Y \approx\frac{g_Y}{4 \pi^5} \,l_w^{-3}\,\left(\frac{R_0}{R_*}\right)^3\mathcal{I}\,(l_{w0}\,m_Y)e^{-m_Y/(\lambda_Y v_\phi)}\,, 
 \label{nucleationcontribution}
\ee
where $g_Y$ is the number of degrees of freedom in field $Y$, $\lambda_Y$ is the coupling between the background field and $Y$, and the dimensionless integral factor $\mathcal{I}(a)$ is 
\bea
\mathcal{I}(a)&\equiv&  \int^\infty_0 dx\, x^2\, \times \frac{\sinh^2\Big[\frac{1}{4} \Big(\sqrt{a^2+x^2}-x\Big)\Big]}{\sinh\Big(\frac{1}{2}\sqrt{a^2+x^2}\Big)\sinh\Big(\frac{1}{2} x\Big)}~~~~~~~(\text{bosons)}\nonumber\\
\mathcal{I}(a)&\equiv&  \int^\infty_0 dx\, x^2\, \times \frac{\cosh\Big(\frac{1}{2}a\Big)-\cosh\Big[\frac{1}{2} \Big(\sqrt{a^2+x^2}-x\Big)\Big]}{2\sinh\Big(\frac{1}{2}\sqrt{a^2+x^2}\Big)\sinh\Big(\frac{1}{2} x\Big)}~~~~~~~(\text{fermions)}
\label{eq:integralfactors}
\eea

The dilution factor $\left(\frac{R_0}{R_*}\right)^3$ in Eq.\,\ref{nucleationcontribution} accounts for the fact that the particles produced within the nucleated bubbles eventually diffuse out over the entire volume of the expanded bubble. Since $R_0\gg R_*$ (recall that $R_0\sim v_\phi^{-1}$ whereas $R_*\sim H^{-1}$), this contribution from bubble nucleation is generally negligible compared to the contribution from subsequent bubble evolution calculated below. Furthermore, note the exponential suppression factor $e^{-m_Y/(\lambda_Y v_\phi)}$: particle $Y$ obtains a contribution to its mass $\Delta m_Y=\lambda_Y v_\phi$ from the phase transition; if this is smaller than the bare mass $m_Y$, the field $Y$ is effectively insensitive to the changing background, hence particle production gets shut off exponentially. Thus, the production of ultraheavy DM $m_\chi\gg \lambda_\chi v_\phi$ during bubble nucleation will be exponentially suppressed.

For a thick-walled bubble, spatial inhomogeneities within the bubble are expected to further suppress particle production compared to the thin-wall case. 

\subsection{Bubble Expansion}

A bubble wall propagating at constant velocity does not produce any particles (for a rigorous derivation, see \cite{Shakya:2023kjf}): one can simply boost to its rest frame, where the configuration is static, hence no particle production can take place. However, in the configurations of interest to us for ultraheavy DM production, bubble walls achieve runaway behavior: they gain the latent energy released from the phase transition and accelerate to larger boost factors as they propagate outwards. Particle production from such accelerating bubble walls can be estimated by making use of the equivalence principle: a nonuniformly accelerating bubble wall is equivalent to a wall at rest in a changing gravitational field, and the familiar calculation of gravitational particle production yields a number density of produced particles $\sim y_\chi^2 \,R_*^{-3}$ \cite{Shakya:2023kjf}. This will also be subdominant to the contribution from bubble collisions discussed in the next subsection.

For thick-wall bubbles, the scalar field might not be at its true minimum anywhere in the bubble when the bubble nucleates, and instead evolves towards the true minimum and performs oscillations around it as the bubble expands. This can also be responsible for some particle production (for related discussions, see \cite{Cutting:2020nla,Ellis:2020nnr}). Since we are focusing on DM particles that are more massive than the background scalar field, such oscillations cannot produce any DM particles. 

\subsection{Bubble Collision}

Particle production from the collision of bubble walls and the subsequent evolution of the background field is a complicated phenomenon due to the highly inhomogeneous nature of the process. The collision of bubbles was first considered in \cite{Hawking:1982ga}, and particle production from such collisions was first studied in detail in \cite{Watkins:1991zt}. Based on the formalism in \cite{Watkins:1991zt}, analytic results were derived in simplified ideal limits in \cite{Falkowski:2012fb}, and recently refined with numerical studies of more realistic setups in \cite{Mansour:2023fwj} and analytic treatment in \cite{Shakya:2023kjf}. Here we provide a brief outline of the formalism; the interested reader is referred to \cite{Watkins:1991zt,Falkowski:2012fb,Mansour:2023fwj,Shakya:2023kjf} for greater details.

The probability of particle production from the dynamics of the field $\phi$ is given by the imaginary part of its effective action, 
\be
\mathcal{P}=2 \,\text{Im}\,(\,\Gamma[\phi\,]\,),
\ee
where $\Gamma[\phi\,]$, the effective action, is the generating functional of one-particle irreducible (1PI) Green
functions  
\be
\Gamma[\phi\,]=\sum_{n=2}^\infty \frac{1}{n !}\int d^4 x_1 ... d^4 x_n \Gamma^{(n)}( x_1,...,x_n)\phi (x_1)...\phi(x_n).
\label{expansion}
\ee
The leading ($n=2$) term suffices for our purposes (we will briefly discuss higher order terms in the next section)
\be
\text{Im}\,(\Gamma[\phi])=\frac{1}{2}\int d^4x_1 d^4 x_2 \phi(x_1)\phi(x_2) \int \frac{d^4 p}{(2\pi)^4}e^{i p (x_1-x_2)} \text{Im}(\tilde{\Gamma}^{(2)}(p^2))\,,
\ee
where $\tilde{\Gamma}^{(2)}$ is the Fourier transform of $\Gamma^{(2)}$. 

The Fourier transform of the background field is $\tilde\phi(p)=\int d^4 x \phi(x) e^{i p x}$. We assume that the bubble walls are planar and collisions occur in the $z-$direction, so that $\tilde\phi(p)=(2\pi)^2 \delta(p_x)\delta(p_y)\tilde \phi(p_z,\omega)$. Using these and the above expressions, the number of particles produced per unit area of colliding bubble walls can be written as \cite{Watkins:1991zt,Falkowski:2012fb}
\be
\frac{N}{A}= 2 \int\frac{dp_z\,d\omega}{(2\pi)^2}\,|\tilde{\phi}(p_z,\omega)|^2 \,\text{Im}[\tilde{\Gamma}^{(2)}(\omega^2-p_z^2)]\,.
\label{interpretation}
\ee

This formula invites the following interpretation. The classical background field configuration can be decomposed via a Fourier transform into its momentum modes. Modes of definite four-momentum $p^2=\omega^2-p_z^2>0$ are to be interpreted as (off-shell) propagating field quanta of the background field with mass $m^2=p^2$ --- we will henceforth denote these as $\phi^*_p$ --- and the probability for each such mode to decay is given by the imaginary part of its Green function.

Following a change of variables, the above formula can be simplified and expressed in terms of the four-momentum of the background field excitations as \cite{Falkowski:2012fb} 
 \be
\frac{N}{A}=\frac{1}{2 \pi^2}\int_{p_{\text{min}}^2}^{p_{\text{max}}^2} d p^2\,f(p^2) \,\text{Im} [\tilde{\Gamma}^{(2)}(p^2)].
\label{number}
\ee
Here $f(p^2)$ encapsulates the details and nature of the collisions as contained in the Fourier decomposition of the background field configuration, representing the \textit{efficiency factor} for particle production at a given energy scale $p$. The integral has a lower limit $p_{min}=2\,m$ (for pair production), set by the mass of the particle species being produced, or the inverse size of the bubble, $(2R_*)^{-1}$ (at lower momenta, the existence of multiple bubbles needs to be taken into account), whichever is greater. The upper cutoff is provided by $p_{max}=2/l_w=2\gamma_w/l_{w0}$, the energy in the two colliding bubble walls, which represents the maximum energy available in the process. The particles produced on the bubble wall collision surface (Eq.\,\ref{number}) will diffuse out over the volume occupied by the bubble, so that the final number density of particles per unit volume is
 \be
n=\frac{3}{4 \pi^2 R_*}\int_{p_{\text{min}}^2}^{p_{\text{max}}^2} d p^2\,f(p^2) \,\text{Im} [\tilde{\Gamma}^{(2)}(p^2)].
\label{number2}
\ee

Similarly, the energy density in particles per unit area is 
 \be
\frac{E}{A}=\frac{1}{2\pi^2}\int_{p_{\text{min}}^2}^{p_{\text{max}}^2} d p^2\,p\,f(p^2) \,\text{Im} [\tilde{\Gamma}^{(2)}(p^2)].
\label{energy}
\ee

The wall collisions can be broadly classified as elastic (where the bubble walls bounce back after collision, restoring the false vacuum in between) or inelastic (where the walls completely dissipate their energy into scalar oscillations, and the true vacuum is established everywhere immediately following the collision). From numerical studies of realistic bubble collision processes, the efficiency factor in the two cases can be parametrized as \cite{Mansour:2023fwj} 
\be
f_{\text{elastic}}(p^2)= f_{\mathrm{PE}}(p^2)+\frac{v_{\phi}^2L_p^2}{15 m_{\mathrm{t}}^2}\exp{\left(\frac{-(p^2 - m_{\mathrm{t}}^2+12 m_{\mathrm{t}}/L_p)^2}{440 \, m_{\mathrm{t}}^2 / L_p^2}\right)}\qquad \text{(elastic collisions)}
\label{eq:elasticfit}
\ee
\be
f_{\text{inelastic}}(p^2)= f_{\mathrm{PE}}(p^2)+\frac{v_{\phi}^2L_p^2}{4 m_{\mathrm{f}}^2}\exp{\left(\frac{-(p^2 - m_{\mathrm{f}}^2+31 m_{\mathrm{f}}/L_p)^2}{650 \, m_{\mathrm{f}}^2 / L_p^2}\right)}\qquad \text{(inelastic collisions)}
\label{eq:inelasticfit}
\ee
Here $m_t,\,m_f$ are the scalar masses in the true and false vacua respectively.  $L_p=\text{min}(R_*, \Gamma_\phi^{-1})$, where $\Gamma_\phi$ is the decay rate of the scalar as it performs oscillations around its true or false minimum and $R_*$ is the typical bubble size at collision, provides a measure of the extent to which scalar oscillations propagate in spacetime. Finally, $f_{\mathrm{PE}}$ is the efficiency factor for a perfectly elastic collision, derived analytically in  \cite{Falkowski:2012fb}
\be
f_{\mathrm{PE}}(p^2)=\frac{16 v_{\phi}^2}{p^4}\, \text{Log}\left[\frac{2(1/l_w)^2-p^2+2(1/l_w)\sqrt{(1/l_w)^2-p^2}}{p^2}\right]\,.
\label{eq:felastic}
\ee
Recall that $l_w=l_{w0}/\gamma_w$ is the Lorentz-contracted bubble wall thickness. 

Note that Eq.\,\ref{eq:elasticfit} and \ref{eq:inelasticfit} contain two distinct contributions: an approximately power law component $f_{\mathrm{PE}}\sim p^{-4}$, originating from the nontrivial dynamics of the background field when the bubbles collide, and an approximately Gaussian peak centered around the mass of the scalar in the relevant vacuum, coming from the oscillation of the scalar field around its relevant minimum after the collision. Since we assume that the DM particle is heavier than the scalar, $m_\chi > m_t,m_f$, the oscillations do not contribute to DM production, and we can ignore the latter component. DM is thus produced solely via  $f(p^2)=f_{\rm PE}$ for both elastic and inelastic collisions.

\subsection{Particle Physics Aspects}

In the formalism above, in Eqs.\,\ref{number2},\,\ref{energy}, the efficiency factor $f(p^2)$ encodes information about the spacetime dynamics of the background field. The particle physics information is encoded in the 2-point 1PI Green function $\Gamma^{(2)}$, to which we now turn our attention.

Using the Optical Theorem, the imaginary part of the 2-point 1PI Green function is given by the sum \cite{Watkins:1991zt,Falkowski:2012fb}
\be
\text{Im} [\tilde{\Gamma}^{(2)}(p^2)]=\frac{1}{2}\sum_k \int d\Pi_k |\bar{\mathcal{M}}(\phi^*_p\to k)|^2
\label{optical}
\ee
Here the sum runs over all possible final states $k$ that can be produced from the background field excitations $\phi^*_p$, $|\bar{\mathcal{M}}(\phi_p^*\to k)|^2$ is the spin-averaged squared amplitude for the decay of $\phi^*_p$ into the given final state $k$, and $d\Pi_k$ denotes the relativistically invariant n-body phase space element. 

Note that the imaginary part of the 2PI Green function is an inclusive quantity that necessitates summing over all possible states $k$ that can contribute. To calculate the overall decay probability of the background field, we therefore need to calculate $|\bar{\mathcal{M}}(\phi^*_p\to k)|^2$ for all particle combinations that are allowed in the setup. However, to calculate the decay probability into a given final state (such as the DM particle), it is sufficient to perform the calculation solely for this channel, and the full sum is not required provided the full decay probability remains smaller than 1, i.e.~that there are no channels that are so strong that particle production backreacts on the system. 

The scalar $\phi$ particles themselves can be produced through the background field excitations, via the quartic term $\frac{\lambda_\phi}{4!}|\phi|^4$ in the scalar potential; this gives rise to $\phi^*_p\to\phi\phi$ (with a single vev insertion) and $\phi^*_p\to 3\phi$ decay processes. These lead to  
\be
\text{Im} [\tilde{\Gamma}^{(2)}(p^2)]_{\phi^*_p\to\phi\phi}=\frac{\lambda_\phi^2 \,v_\phi^2}{8\pi}  (1-4m_\phi^2/p^2)\,\Theta(p-2m_\phi)
\label{2scalar}
\ee
and
\be
\text{Im} [\tilde{\Gamma}^{(2)}(p^2)]_{\phi^*_p\to3\phi}=\frac{\lambda_\phi^2 \,p^2}{3072\,\pi^3}  (1-9m_\phi^2/p^2)\,\Theta(p-3m_\phi)
\label{3scalar}
\ee
Note that the three-body process is suppressed relative to the two-body process by a loop factor due to an additional particle in the final state, but is proportional to $p^2$ rather than $v_\phi^2$, hence can become more important at higher $p^2$ as it can be realized even in the $v_\phi\to 0$ limit where the symmetry is unbroken. 

For scalar DM, which couples as $\frac{\lambda_s}{4} \phi^2\chi_s^2$, the formulae for two- and three-body decays $\phi^*_p\to\chi_s\chi_s$ and $\phi^*_p\to\phi\chi_s\chi_s$ are analogous to Eqs.\,\ref{2scalar},\,\ref{3scalar}, with $\lambda_\phi\to\lambda_s$, $3072\to 1024$ due to modified symmetry factors, and appropriate modifications of the final state masses in the phase space factors and step functions.

For fermion DM, the relevant expression is
\be
\text{Im} [\tilde{\Gamma}^{(2)}(p^2)]_{\phi^*_p\to \chi_ f\bar{\chi}_f}=\frac{y_f^2}{8\pi} p^2 (1-4m_{\chi_f}^2/p^2)^{3/2}\,\Theta(p^2-4m_{\chi_f}^2)\,.
\label{fermion}
\ee
Note that this quantity is proportional to $p^2$ and can occur in the $v_\phi\to 0$ limit of unbroken symmetry, similar to the three-body scalar decay channel above.

The calculation for vector DM, and final states involving gauge bosons in general, is more subtle and requires a discussion of the gauge dependence of the formalism. This will be the subject of the next section.

\section{Gauge Dependence and Production of Gauge Bosons}
\label{sec:gaugedependence}

Here, we consider the case where the scalar vev breaks a local symmetry, and discuss the production of the massive gauge boson $V$ associated with the broken symmetry.  The results can be extended in a straightforward manner to other vector bosons, in particular the vector DM candidate we are interested in. 

\subsection{Gauge Dependence}
\label{gauged}

To understand the subtleties regarding the gauge dependence of the formalism, let us consider the decay of a background field excitation into two gauge bosons, $\phi^*_p\to V V$, which occurs via the interaction term $g v_\phi \phi V_\mu V^\mu$. The calculation of the squared amplitude of this process requires a sum over the gauge boson polarizations. Its general form, in $R_\xi$ gauge, is 
\be
\sum \epsilon^\mu\epsilon^\nu\to -g^{\mu\nu}+(1-\xi)\frac{p^\mu p^\nu}{p^2-\xi m_V^2}\,.
\label{polarizationsum}
\ee
 Recall that in $R_\xi$ gauge, one must also add the contributions from the Goldstone and ghost fields, which have mass $m^2=\xi m_V^2$. For a physical process, the choice of $\xi$ and the separation of the degrees of freedom into gauge, Goldstone, and ghost fields is simply a matter of bookkeeping, and the final result should be gauge-invariant, i.e.\,$\xi-$independent. As we will see below, this will not be the case for the above configuration and formalism describing bubble collisions, hence greater care is needed to avoid spurious results.  

Generally, a convenient choice is unitary gauge $(\xi\to\infty)$, where the Goldstone and ghost fields decouple, and one simply needs to consider the gauge degrees of freedom, for which the above sum over polarization reduces to the familiar expression $\sum \epsilon^\mu\epsilon^\nu\to -g^{\mu\nu}+\frac{p^\mu p^\nu}{m_V^2}$. Using this, the squared amplitude for the  $\phi^*_p\to V V$ process can be calculated to be 
\be
|\bar{\mathcal{M}}(\phi^*_p\to V V)|^2=g^2 m_V^2 \left(3-\frac{p^2}{m_V^2}+\frac{p^4}{4 m_V^4}\right)\,~~~~~\text{(Unitary gauge)}.
\label{MgaugeU}
\ee

One can, instead, perform this calculation in Feynman-'t Hooft gauge $(\xi=1)$. With this choice, the polarization sum yields $\sum \epsilon^\mu\epsilon^\nu\to -g^{\mu\nu}$, and one has to add the Goldstone and ghost contributions separately. Adding these contributions together results in the following expression for the squared amplitude
\be
|\bar{\mathcal{M}}(\phi_p^*\to V V)|^2=g^2 m_V^2 \left(3-\frac{p^2}{m_V^2}+\frac{\lambda_\phi^2}{g^4}\right)\,~~~~~\text{(Feynman-'t Hooft gauge)}.
\label{MgaugeF}
\ee

For a physical process, both results should match and give the correct (physical) result. When the decaying mode corresponds to an on-shell $\phi$ particle, i.e.~$p^2=m_\phi^2$, this is indeed seen to be true: in this case $p^4/(4m_V^4)=m_\phi^4/(4m_V^4)=\lambda_\phi^2/g^4$, hence the final expressions in the parentheses in the two equations are identical. The problem arises when the excitation $\phi_p^*$ is taken off-shell, i.e.~$p^2\neq m_\phi^2$. In this case, the two expressions clearly disagree: in particular, at large $p^2\gg m_\phi^2, m_V^2$, the unitary gauge result scales as $\sim g^2\,p^4/m_V^2$\,, whereas the Feynman-'t Hooft gauge result scales as $\sim-g^2\,p^2$. Clearly, this discrepancy persists even after the sum over modes (Eq.\,\ref{number}) is performed; hence the final result for the number density of gauge bosons produced from a bubble collision appears to be gauge-dependent. 

Both results above, Eq.\,\ref{MgaugeU} and Eq.\,\ref{MgaugeF}, are however unphysical. The Feynman-'t Hooft gauge result gives a negative decay probability at large $p^2$, which is clearly unphysical. The problem with the unitary gauge result can be seen most clearly by considering the analogous contribution from the higher multiplicity process $\phi^*_p\to 4V$. Compared to the $\phi^*_p\to 2V$ process, the $4V$ process has an additional scalar propagator, whose contribution to the amplitude squared scales approximately as $\sim \frac{1}{p^4}$; two additional vector bosons in the final state, which give additional phase space factors $(\frac{d^3 k}{(2\pi)^3 2 p})^2\sim (\frac{p^2}{4\pi^2})^2$; and a sum over the two additional gauge boson polarization vectors, which yields another factor of $\left(3-\frac{p^2}{m_V^2}+\frac{p^4}{4 m_V^4}\right)$. Thus, we can estimate the leading order contributions at large $p$ from the $\phi^*_p\to4V$ and $\phi^*_p\to 2V$ processes to the imaginary part of the two point 1PI Green function in unitary gauge to be
\be
\text{Im} [\tilde{\Gamma}^{(2)}(p^2)]_{\phi^*_p\to 4V}\sim\frac{g^6 m_V^2}{8\pi (4\pi^2)^2} \left(\frac{p^4}{4 m_V^4}\right)^2 \,,~~~~~
\text{Im} [\tilde{\Gamma}^{(2)}(p^2)]_{\phi^*_p\to 2V}\sim\frac{g^2 m_V^2}{8\pi} \left(\frac{p^4}{4 m_V^4}\right)\,. 
\ee
Therefore, the $\phi^*_p\to 4V$ contribution appears to grow faster than the $\phi^*_p\to 2V$ contribution at large $p^2$. By similar arguments, processes with higher vector boson multiplicity in the final state should grow even faster with $p^2$. If true, this would preclude the calculation of Eq.\,\ref{optical}, which is an inclusive quantity that requires the addition of all of these higher order processes. More worryingly, this unabated growth suggests a breakdown of perturbativity despite the absence of any strong coupling in the theory. This is a clear indication that the growth of the squared amplitude with energy in Eq.\,\ref{MgaugeU} is spurious. 

A general form worth noting is
\be
\text{Im} [\tilde{\Gamma}^{(2)}(p^2)]_{\phi^*_p\to 2V}=g^2 m_V^2 \left(2+\frac{(p^2-2m_V^2)^2}{4 m_V^4}\right)\sqrt{1-\frac{4m_V^2}{p^2}}+\frac{g^2}{4 m_V^2}(m_\phi^4-p^4)\sqrt{1-\frac{4\xi m_V^2}{p^2}}\,.
\ee

Since the gauge, Goldstone, and ghost fields have unequal masses in general $R_\xi$ gauge, this leads to unequal phase space weights for any finite $p$, so that the sum of their contributions cannot be expressed as a single squared matrix element as in Eqs.\,\ref{MgaugeU} and \ref{MgaugeF}. Note that the above two cases are the only exceptions to this: in Feynman-'t Hooft gauge these masses are equal, so that the phase space factor is the same for all contributions a nd can be factored out, whereas in unitary gauge the Goldstone and ghost fields decouple, and only the gauge component contributes to the amplitude. One can also write down the following asymptotic expansion:
\be
|\bar{\mathcal{M}}(\phi_p^*\to V V)|^2=g^2 m_V^2 \times \left\{
\begin{array}{ll}
\frac{(\xi-3) p^2}{2 m_V^2}+\frac{\lambda_\phi^2}{g^4}+3& ~{\rm for}~ \frac{p^2}{m_V^2} \gg \xi ,1  \\
 \frac{p^4}{4m_V^4}-\frac{p^2}{m_V^2} +3& ~{\rm for}~ \xi \gg \frac{p^2}{m_V^2}  ,1
\end{array} \right.
~~\text{($R_\xi$ gauge).}
\label{MgaugeR}
\ee
This further illustrates that the high-energy behavior of the off-shell $\phi_p^*$ decay squared amplitude is gauge-dependent and can become ${\cal O} (g^2 p^4/m_V^2)$, ${\cal O} (g^2 p^2)$, or ${\cal O} (g^2 m_V^2)$, depending on the value of $\xi$. In particular, in the Fried-Yennie gauge ($\xi = 3$), both the $p^4$ and $p^2$ terms are absent in the large-$p^2$ expansion.

In any gauge-specific calculation, the problem arises due to the inclusion of unphysical contributions that do not get cancelled. For the unitary gauge result, note that the problematic final term in Eq.\,\ref{MgaugeU} comes from the prescription of taking $\epsilon_L\to p_i/m_V$ for the production of two longitudinal modes. However, in the large $p_i$ limit, we know that the emission of the longitudinal component of the gauge boson should be equivalent to the emission of the corresponding Goldstone boson ``eaten" by the gauge boson, as prescribed by the Goldstone Equivalence Theorem (GET). Since the Goldstone is a component of the scalar field, this contribution to the matrix element should therefore scale as $\sim\lambda_\phi^2 v_\phi^2$ at high energies, and this $\sim p^4/m_V^2$ growth is unphysical. The Feynman-'t Hooft gauge result rectifies this problem: in Eq.\,\ref{MgaugeF}, the final term, which comes from adding the emission of two Goldstone bosons, indeed scales as $\sim\lambda_\phi^2 v_\phi^2$ rather than $\sim p^4/m_V^2$, hence the spurious growth with energy encountered in the unitarty gauge calculation is eliminated and the behavior anticipated from the GET is recovered.\,\footnote{For a related discussion of an equivalent gauge that makes the Goldstone equivalence manifest at high energies, see \cite{Cuomo:2019siu,Wulzer:2013mza}.} However, the polarization sum $\sum \epsilon^\mu\epsilon^\nu\to -g^{\mu\nu}$, which includes physical as well as unphysical contributions, now gives rise to the negative (second) term in Eq.\,\ref{MgaugeF}, resulting in unphysical (negative) probabilities, suggesting that unphysical contributions to the polarization sum have not been cancelled in the final result.

Fully restoring the gauge independence of the calculation requires choosing an initial configuration that is physical, which should result in the cancellation of all unphysical contributions and guarantee gauge independence of the final result. 
The gauge dependence of the formalism we are considering here can be traced to the assumption that the Fourier transform of the classical field configuration can be interpreted as a collection of off-shell field quanta of different effective masses corresponding to different four-momenta (see Eq.\,\ref{interpretation} and the paragraph below it). Since an ensemble of off-shell quanta is not a physical configuration, there is no guarantee that the ensuing calculation is gauge invariant.

The issue at hand can be understood in analogy with the familiar example of gauge boson scattering, $VV\to VV$, at center of mass energy $E$. If one only considers the process $VV\to \phi^*\to VV$ mediated by an s-channel scalar particle, the leading contribution grows as $\sim E^4$. As is well known, this term is cancelled when adding all other diagrams that contribute to $VV\to VV$ scattering; however, to obtain this physical result, it is necessary to sum over all contributions that are relevant. Similarly, the spurious pieces in Eq.\,\ref{MgaugeU} and Eq.\,\ref{MgaugeF} should also be similarly cancelled if all contributions relevant to the physical process at hand are appropriately included. However, our starting point for the calculation is not a physical process (as in $VV$ scattering) but a collection of off-shell massive excitations $\phi^*_p$ (akin to only picking out the $VV\to \phi^*\to VV$ contribution for vector boson scattering, which is incomplete), which cannot ensure gauge invariance. \footnote{It should be noted that there exist various techniques to address similar gauge invariance issues in other contexts,  see e.g. \cite{Papavassiliou:1997fn,Papavassiliou:1997pb,Duch:2018ucs}; while they might lead to some simplifications in our calculations, they will not completely solve the problem since we are considering decays of unphysical off-shell field excitations.}

This suggests that the decomposition of the classical scalar field configuration at bubble collision into a collection of Fourier modes of off-shell field quanta in the above formalism, and considering only the leading order ($n=2$) terms in Eq.\,\ref{expansion}, misses contributions that are relevant. It is not clear what these missing ingredients are, but there are likely several things that might be relevant.  Including the higher order terms in the expansion in Eq.\,\ref{expansion} is certainly necessary. Other known techniques, such as gradient expansion or dimensional reduction, might also provide some insight towards a resolution of the problem. Depending on the gauge of choice, other fields might develop profiles and contribute to the bubble walls in addition to the scalar field. Likewise, in the above discussions we have used matrix elements corresponding to the theory in the true vacuum, but the rigorous construction of an S-matrix element for the decay of a transient excitation across two stable points of a theory likely involves more subtleties that would need to be addressed.

Without knowing all of the relevant contributions, a fully gauge invariant calculation cannot be performed. Nevertheless, as we will see below, it is still possible to extract meaningful physical, gauge independent results from the known contributions by making use of the Goldstone Equivalence Theorem, which provides a practical path to performing the necessary calculation. 

\subsection{High Energy Behavior}

Practically speaking, the spurious results above arise from unphysical terms in the sum over gauge boson polarizations. In unitary gauge $(\xi\to\infty)$ (Eq.\,\ref{MgaugeU}), the third term contributes the $p^4/m_V^2$ term that is unphysical and should have been cancelled by contributions from other relevant diagrams. On the other hand, in Feynman-'t Hooft gauge, $\sum \epsilon^\mu\epsilon^\nu\to -g^{\mu\nu}$ sums over all polarizations, \textit{including} unphysical ones; these, again, should be cancelled by contributions from other relevant diagrams, but remain in their absence and contribute the unphysical $-p^2$ piece in Eq.\,\ref{MgaugeF}. Therefore, a practical solution would be to only pick out physical contributions from physically allowed polarization states explicitly when performing the sum. 

Instead of using Eq.\,\ref{polarizationsum} to perform the sum over polarizations, we can instead explicitly pick the polarization states. For a gauge boson moving in the $z-$direction, the transverse (T) polarization states are $\epsilon_T^\mu=(0,1,0,0),~(0,0,1,0)$, whereas the longitudinal (L) polarization vector is $ \epsilon_L^\mu=(p/m_V,0,0,E_V/m_V)$. The latter has the problematic $p/m_V$ growth at large $p$; however, this can be tamed with the Goldstone Equivalence Theorem (GET), which states that at high energies the amplitude for the emission of a longitudinally polarized massive gauge boson becomes equal to the amplitude for emission of the Goldstone mode $\phi_G$ ``eaten" by the gauge boson, up to corrections of order $\mathcal{O}(m_V^2/p^2)$. Thus, even in the absence of all contributing diagrams, the GET provides a prescription for extracting the physical behavior of the longitudinal mode at high energies that is free of unphysical contributions and does not require choosing a specific gauge for the calculation. 

We can apply this strategy to the $\phi^*_p\to V V$ process discussed above to extract its high energy behavior. Three polarization combinations contribute to the calculation of $|\mathcal{M}|^2$ :
\begin{itemize}
\item TT: The emission of transverse modes is well behaved, and gives $2m_V^2$. 
\item LL: Using the GET, this is equivalent to the emission of two Goldstones $\phi^*_p\to\phi_G\phi_G$, and gives  $\lambda^2_\phi v_\phi^2$, or equivalently $(\lambda^2_\phi/g^2)m_V^2$.
\item TL: Invoking the GET, we need to calculate $\phi^*_p\to V_T(p_1) \phi_G(p_2)$ to obtain the high energy behavior of this contribution. This diagram comes from the kinetic term of the scalar, and has a vertex factor $i g (p^\mu+p_2^\mu)$ that contracts with the gauge boson polarization $\epsilon_T^\mu$. In the rest frame of $\phi^*_p$, the two emitted particles are back to back, these vectors are othogonal, and this contraction vanishes, hence this combination does not contribute at high energies.\,\footnote{Strictly speaking, the collection of background field excitation modes has a distribution of $p_z$, and there is no frame where they are all collectively at rest. Nevertheless, we have assumed that for the decay of each excitation, the calculation can be performed in its rest frame, as is conventionally done for a collection of particles with a distribution of momenta, otherwise the result is not Lorentz invariant.}
\end{itemize}

Adding these contributions, we obtain the following form of the squared amplitude at high energies:
\be
|\bar{\mathcal{M}}(\phi^*_p\to V V)|^2\xrightarrow{p^2>m_V^2}(2 g^2+\frac{\lambda_\phi^2}{g^2})\, m_V^2\, (1+\mathcal{O}(m_V^2/p^2))\,~~~~~\text{(Goldstone Equivalence Theorem)}.
\label{GETresult}
\ee
Note that this result is well-behaved and contains neither the spurious $\propto p^4$ growing term from the unitary gauge calculation nor the $-p^2$ term from the calculation in Feynman-'t Hooft gauge; the above prescription has eliminated all unphysical ingredients and picked out the relevant physical contributions from the process at hand, without requiring any explicit computation in a specific gauge. This will continue to be the case for all other relevant diagrams, as we discuss below. Therefore, we can interpolate between the low energy behavior (Eq.\,\ref{MgaugeU}) and the high energy behavior (Eq.\,\ref{GETresult}) to obtain an approximate result for the production of gauge bosons; this method introduces inaccuracies in the intermediate regime ($p^2\sim m_V\sim v_\phi$), but the final result for the total number of particles is expected to be correct within an $\mathcal{O}(1)$ factor.

\subsection{Other Processes}
\label{others}

In addition to $\phi^*_p\to VV$, there also exists the three-body decay process $\phi^*_p \to \phi V V$. Naive gauge-specific calculations also give unphysical results for this decay channel for the reasons described above, but one can similarly use the prescription above to estimate its high energy behavior:
\be
\phi^*_p \to \phi V V: ~~~~~ |\bar{\mathcal{M}}|^2_{~p\,<\,v_\phi}\sim g^4 \left(3-\frac{p^2}{m_V^2}+\frac{p^4}{4 m_V^4}\right),~~~~~  |\bar{\mathcal{M}}|^2_{~p\,>\,v_\phi}\sim \lambda_\phi^2+2 g^4 \,.
\label{3sg}
\ee
Note that this three-body decay will be phase-space suppressed relative to the two-body decay by a factor $\sim (16\pi^2)^{-1}$ due to an additional particle in the final state but can nevertheless dominate at large $p$, analogously to the two and three-body scalar decay processes in Eqs.\ref{2scalar},\,\ref{3scalar}.

Similarly, particles can also be produced due to interactions between multiple Fourier modes, e.g.\,$\phi^*_{p1}\phi^*_{p2} \to \phi \phi, VV, \chi\chi$.  These correspond to higher order terms in the expansion in Eq.\,\ref{expansion}. The Fourier transform of the additional $\phi^*_{p_i}$ excitation in the initial state scales as $\sim (v_\phi/p^2)^2$, hence  the higher order $\phi^*_{p1}\phi^*_{p2}\to \phi \phi, VV, \chi\chi$ processes are subdominant for very heavy DM but can introduce $O(1)$ corrections for DM whose  mass arises from its coupling to $\phi$. 

However, further higher order terms corresponding to additional $\phi^*$ in the initial configuration, or additional vev insertions, can be important in processes involving particles with mass lighter than the scale of symmetry breaking. For concreteness, consider the scalar DM candidate with $\chi_s$ with mass $m_{\chi_s}$ that couples to $\phi$ via the interaction $\frac{\lambda_s}{4}\phi^2\chi_s^2$. A double $\phi^*$ insertion on the DM state introduces a factor $\sim\left(\frac{v_\phi}{ p^2}\right)^4$ from the Fourier transform of the additional $\phi$ excitations, vertex factor $\lambda_s^2$, an additional DM propagator, which gives a $\frac{1}{p^4}$ contribution, and phase space factors that scale with some appropriate power of $p$, resulting in an overall contribution that is a factor $\sim\lambda_s^2\left(\frac{v_\phi}{ p}\right)^4$ larger than the original diagram. Therefore, one cannot truncate the expansion in Eq.\,\ref{expansion} at the leading term if $p^2 \lesssim \lambda_s\,v_\phi^2$. However, as mentioned in Sec.\,\ref{subsec:content}, we restrict ourselves to $m_{\chi_s}^2 > \frac{1}{2}\lambda_s v_\phi^2$, hence it is consistent to ignore such higher order corrections in this region of parameter space.

\subsection{Backreaction Effects}

In the previous subsections, we have highlighted two important aspects of particle production from bubble dynamics: (i) the calculation for gauge boson production is gauge dependent, and the correct scaling at high energies can be obtained by making use of the Goldstone equivalence theorem; (ii) at high energies, the emission of three (scalar or gauge) bosons is enhanced compared to the emission of two bosons despite the phase space suppression, as the squared matrix element scales as $\sim p^2$ in the former case and as $\sim v_\phi^2$ in the latter. Here we briefly discuss the relevance of these results for backreaction effects on bubble dynamics and DM abundance; for more detailed discussions on backreaction effects, see \cite{Shakya:2023kjf}.

If the energy density in the produced particles is a significant fraction of the latent energy released in the phase transition, this creates a backreaction effect on the bubble dynamics, which should be appropriately taken into account for phenomenological applications such as the calculation of gravitational waves from the scalar field at and after bubble collision, and for calculating the relic abundance of DM. From the formula for the energy density in the produced particles (Eq.\,\ref{energy}), we see that the energy density depends critically on the form of $\text{Im} [\tilde{\Gamma}^{(2)}]$, or equivalently the matrix element $|\mathcal{M}|^2$. Since $f(p^2)\sim p^{-4}$ (Eq.\,\ref{eq:felastic}),  if $\text{Im} [\tilde{\Gamma}^{(2)}]\sim p^{x}$ with $x>1$, the energy density in particles grows as a positive power of $p^2$, hence backreaction can become significant for large values of $p$. 

Previous works  \cite{Watkins:1991zt,Falkowski:2012fb} concluded that the production of scalars (for which the matrix element for pair production scales as $|\mathcal{M}|^2\propto p^0$) is not strong enough, but pair production of fermions ($|\mathcal{M}|^2\propto p^2$) and gauge bosons  ($|\mathcal{M}|^2\propto p^4$) can be efficient enough to backreact on bubble dynamics; in particular, the production of gauge bosons is so efficient at high energies (as can be seen from inserting $|\mathcal{M}|^2\propto p^4$ in Eq.\,\ref{energy}) that it significantly reduces the energy available for other states, so that it precludes the possibility for scalar DM in a first-order electroweak phase transition (and other gauged transitions in general), whereas fermion DM remains marginally possible, and vector DM can be realized across a large range of masses from $\sim 1-10^8$ TeV. 

Our results disagree with these conclusions. As discussed in Sec \ref{gauged} above, the $|\mathcal{M}|^2\propto p^4$ scaling for gauge boson pair production at large $p^2$ is a spurious gauge artifact, and the correct scaling at energies above the scale of symmetry breaking is in fact $|\mathcal{M}|^2\propto p^0$ (Eq.\,\ref{GETresult}). However, we noted that the three-body decay processes involving scalars and gauge bosons $\phi^*_p\to 3 \phi,~\phi VV$, Eq.\,\ref{3scalar},\,\ref{3sg} (which were not considered in \cite{Watkins:1991zt,Falkowski:2012fb}, but discussed in \cite{Shakya:2023kjf}), do scale as $\text{Im} [\tilde{\Gamma}^{(2)}]\propto p^2$ (albeit with additional phase space suppression). As a result, there is no process with $\text{Im} [\tilde{\Gamma}^{(2)}]\propto p^4$ that can backreact severely on the bubble dynamics, but the processes with $\text{Im} [\tilde{\Gamma}^{(2)}]\propto p^2$ can backreact if the associated coupling is sufficiently large; see \cite{Shakya:2023kjf} for a more detailed and qualitative treatment. As we will see below, these modified results open up significant parameter space for scalar, fermion, and vector DM.

\section{Dark Matter Production}
\label{sec:dmproduction}

Before exploring the parameter space where the above mechanism yields the correct dark matter relic abundance, we first discuss other DM production mechanisms that might be active at various stages of the phase transition in different cases. Here we will use the general interaction form $\frac{1}{4}\lambda_\chi \phi^2\chi^2$, using the general notation $\chi$ for DM and $\lambda_\chi$ for its coupling to the background field, as the discussion is broadly applicable to DM of arbitrary spin. We will revert to spin-specific notations as introduced in Sec.\,\ref{subsec:content} where this is not the case. Here we are simply interested in obtaining order-of-magnitude estimates, hence we will make use of several approximations without worrying about $\mathcal{O}(1)$ factors. 

The relic abundance of DM can be written as
\be
\Omega_\chi h^2=6.3\times 10^8 \frac{m_\chi}{\text{GeV}}\frac{n_\chi}{g_{*}(T_*) T_*^3}\,,
\label{dmgeneral}
\ee
where $n_\chi$ is the DM number density at the time of production (in our case, given by Eq.\,\ref{number2}), when the temperature of the thermal bath after the FOPT is $T_*$, and $g_{*}$ is the number of degrees of freedom in the bath at this time. Recall that the observed abundance of DM corresponds to $\Omega_{DM} h^2=0.12$. 

\subsection{Pre-Transition Contributions}

The early Universe before the phase transition could already contain some DM abundance. If DM is in thermal equilibrium with the bath, it obtains an equilibrium number density $n_\chi \sim T^3$ before undergoing thermal freezeout, during which its abundance can drop exponentially for $T<m_\chi$. For DM masses beyond the unitarity bound $\mathcal {O}(100)$ TeV, the frozen-out relic abundance is too large and not viable; in this case, the abundance can be suppressed below $\Omega_{DM} h^2=0.12$ if there is a large amount of entropy injection that dilutes the DM yield by several orders of magnitude. This can occur if the transition is supercooled, or through late decays of some heavy particles, or if the dark sector is decoupled from the visible (SM) sector and colder. 

\vskip 1 cm

\noindent \textit{Freeze-in:}

Another viable possibility for masses beyond the unitarity bound involves DM never reaching equilibrium with the dark or SM bath, but only realizing smaller, nonthermal abundances via the freeze-in mechanism \cite{McDonald:2001vt,Hall:2009bx}. This can occur if the reheat temperature of the Universe $T_{R}$, defined as the maximum temperature of the thermal bath after the onset of radiation domination following inflation, is below the freezeout temperature for DM.\footnote{The Universe could have reached temperatures higher than $T_R$ during the thermalization phase between the end of inflation and the onset of radiation domination \cite{Giudice:2000ex}, which can also enable the production of massive particles \cite{Chung:1998rq}.} Alternately, non-equilibrium is maintained at higher temperatures above the DM mass if the associated coupling is sufficiently small; this is achieved for $\lambda_\chi \lesssim \sqrt{v_\phi/M_{Pl}}$. In both cases, DM is produced gradually via freeze-in processes such as $\phi\phi\to \chi\chi$. 

If $\phi$ is in thermal equilibrium with the SM bath and the $\phi\phi\to \chi\chi$ process originates from a renormalizable coupling $\frac{\lambda_\chi}{4} \phi^2\chi^2$, as would be the case if DM is a scalar or a vector, the DM relic abundance from this contribution is \cite{Hall:2009bx}
\be
\Omega_\chi h^2 \sim 10^{20}\, \lambda_\chi^2 e^{-2 m_\chi/T_R}\,\left(\frac{T_n}{T_*}\right)^3,
\label{renormalizableIR}
\ee
where the exponential factor has been added to account for the Boltzmann suppression that exists if $m_\chi>T_R$, and the $\left(\frac{T_n}{T_*}\right)^3$ factor accounts for entropy dilution from the energy injection from the FOPT. Recall that we have assumed that none of the dark sector particles can decay into DM, so that there is no contribution from $\phi\to\chi\chi$ or other dark sector decays.

If, instead, this annihilation occurs through a higher dimensional operator of the form $\frac{1}{\Lambda}\phi\phi \chi \chi$, as could be the case for fermion DM, or for DM in the presence of a heavy mediator (in this case additional considerations might be relevant, see \cite{Frangipane:2021rtf}), the abundance is UV-dominated  \cite{Elahi:2014fsa,Roland:2016gli}, i.e.~receives dominant contributions at the largest temperatures. If $\phi$ is in equilibrium with the SM bath, the UV freeze-in abundance of fermion DM produced from this operator is \cite{Roland:2016gli}
\be
\Omega_\chi h^2\sim 0.1 \left(\frac{m_\chi}{\text{GeV}}\right)\left(\frac{1000\, T_R\,M_{Pl}}{\Lambda^2}\right) e^{-2 m_\chi/T_R}\left(\frac{T_n}{T_*}\right)^3\,.
\ee 
 If $\phi$ is out of equilibrium with the SM bath, then $\phi$ itself gets produced via freeze-in processes, and its subsequent annihilations produce DM. The DM abundance in this case can be calculated analogously using appropriately modified versions of the above formulae. 

Note that the above contributions only exist in the presence of a dark sector bath (Scenarios I, II in Section \ref{subsec:runaway}), but are irrelevant in scenarios where a dark sector bath is essentially absent (Scenarios III, IV).

\subsection{Other Contributions during the Transition}

\noindent \textit{Wall-plasma interactions:}

In the presence of a dark sector bath, additional DM production can occur when $\phi$ particles present in the bath interact with relativistic bubble walls as they cross into true vacuum bubbles\,\footnote{It is also worth noting here that for properly calculating the effects of particles transitioning across the bubble wall in a gauged theory, the fields need to be appropriately quantized across the bubble wall, and the quantization of the longitudinal mode of the gauge boson in particular is subtle \cite{Azatov:2023xem}.} \cite{Azatov:2020ufh,Azatov:2021ifm,Azatov:2022tii,Baldes:2022oev}. For a renormalizable interaction of the form $\frac{\lambda_\chi}{4} \phi^2\chi^2$, the probability for a $\phi$ particle to up-scatter into $\chi\chi$ as it transitions across a bubble wall is \cite{Azatov:2021ifm}
\be
P(\phi\to\chi\chi)=\frac{\lambda_\chi^2\,v_\phi^2}{96\pi^2m_\chi^2}\,.
\ee
Here, $\lambda_\chi^2 v_\phi^2/m_\chi^2$ can be thought of as an effective mixing angle between $\phi$ and the $\chi\chi$ state. This transition requires the crossing of the $\phi$ particle across the bubble wall to be non-adiabatic, i.e.\,the evolution occurs sufficiently rapidly that the $\phi$ particle cannot adiabatically track the massive $\phi$ eigenstate across the wall but instead upscatters into the $\chi\chi$ combination. This non-adiabaticity condition is given by\cite{Azatov:2020ufh,Azatov:2021ifm,Baldes:2022oev}
\be
\gamma_w>\frac{l_{w0}\, m_\chi^2}{T_*}\sim \frac{m_\chi^2}{v_\phi^2}\,,
\label{nonadiabatic}
\ee 
where in the second step we have assumed $l_{w0}^{-1}\sim T_*\sim v_\phi$. Thus we see that merely being above the kinematic threshold for DM production, $\gamma_w T_*> m_\chi$, is not sufficient; the non-adiabaticity condition requires $\gamma_w$ to be larger than this by an additional factor of $m_\chi/v_\phi$. Provided Eq.\,\ref{nonadiabatic} is satisfied, the DM contribution from the above particle-bubble interaction process is \cite{Azatov:2021ifm}
\be
\Omega_\chi h^2\approx\frac{1.35\times10^5 \lambda_\chi^2}{g_{*}}\frac{v_\phi^2}{m_\chi^2}\frac{m_\chi}{\text{GeV}}\left(\frac{T_n}{T_*}\right)^3.
\label{eq:nonadiabatic}
\ee

\noindent \textit{Bubbletron:}

Particle-bubble interactions can also produce DM via another mechanism. Particles gaining mass from bubble crossing  can produce shells of accelerated particles with large boost factors that get dragged along with the bubble walls, and when the bubble walls collide, these particle shells also collide with high energies, 
an event dubbed a ``bubbletron" \cite{Baldes:2023fsp,Baldes:2024wuz}. The realization of this configuration requires the particles to retain their energies over the course of the expansion phase (i.e.\,not interact with other particles in their vicinity). In this case, since the accelerated particles gain boost factors comparable to the boost factor of the wall $\sim\gamma_w$, their collisions can also produce very heavy DM. The modeling of such particles shells and their phase space distributions and collisions is complicated and the subject of ongoing work in the literature (see \cite{Baldes:2023fsp}). Here, using the simple estimates from \cite{Baldes:2023fsp} (see also \cite{Baldes:2024wuz} for more recent detailed calculations), we approximate the DM contribution from this process to be
\be
\Omega_\chi h^2\sim 10^{-15}\, \lambda_\chi^2 \,\frac{\beta}{H}\left(\frac{T_n}{T_*}\right)^4 \frac{m_\chi\,v_\phi}{\text{GeV}^2}\,.
\label{eq:bubbletron}
\ee 
We have checked that this is consistent with the DM abundance estimated by (optimistically) assuming that a collection of particles with a thermal abundance $\sim T_n^3$ undergoes collisions with energy $\sim \gamma_w/l_{w0}$ for a duration $\sim R_*^{-1}$.

\subsection{Post-Transition Contributions}

The phase transition completes when the bubbles of true vacuum collide, and the energy carried by the bubble walls dissipates into dark sector particles and scalar waves. Recall that the bubble collisions produce particles with very high energies, up to $\sim 1/l_w=\gamma_w/l_{w0}\sim \gamma_w v_\phi$. Therefore, these particles are energetic enough to produce DM through their collisions even when the DM mass is significantly higher than the temperature of the bath $T_*$ or the scale of symmetry breaking $v_\phi$ (or even the reheat temperature $T_R$). Note that this post-transition contribution exists for all FOPTs (Scenarios I-IV in Sec.\,\ref{subsec:runaway}), including the ones that are devoid of a dark sector bath before the phase transition, since the bubble collisions populate dark sector particles in all cases. 

We can make some simple qualitative observations to estimate the importance of this effect. From the previous sections, we know that the leading order diagram for the production of a particle of any spin scales as $|\mathcal{M}|^2\propto p^2$ at high energies through either two- or three-body decays; therefore, the abundances of dark sector particles at a given energy are approximately proportional to their couplings to the background field. From this, it is straightforward to deduce that any dark sector particle with a coupling to the background field smaller than that of the DM particle is produced with lower abundance than DM (beyond the kinematic threshold where DM can be produced), and cannot affect its abundance. It is only possible to substantially alter the DM abundance in the presence of a dark sector particle that has a larger coupling to the background field than the DM particle. 

\noindent \textit{Scalar Annihilation:}

The existence of such particles is a model-dependent question; nevertheless, in the minimal model we can consider the contribution from the production and subsequent annihilations of the scalar field $\phi$ itself. In the presence of a large self coupling ($\lambda_\phi>\lambda_\chi$), $\phi$ particles are produced at high energies with a greater abundance than $\chi$ particles through the $\phi^*_p\to 3\phi$ process (Eq.\,\ref{3scalar}). These high energy $\phi$ particles decay with a finite lifetime into the SM bath, but before they decay, they can self-scatter and thermalize through the quartic coupling, approaching a thermal distribution. During this thermalization process, they can also (with a lower probability) annihilate into DM states. Since we are interested in $m_\phi\ll m_\chi$, we can only consider the fraction of the $\phi$ population with energies greater than the DM mass. For this population, a simple estimate for the number density of DM particles from $\phi$ annihilation in a Hubble time after the completion of the phase transition is
\be
n_{\chi\,(\phi\phi\to\chi\chi)}\approx n_{\phi\,(E_\phi>2m_\chi)} n_\phi \,\sigma_{\phi\phi\to\chi\chi}/H\approx  \frac{\lambda_\chi^2}{32\pi \sqrt{g_*} }\frac{n_{\phi\,(E_\phi>2m_\chi)} n_\phi\, M_{Pl}}{m_\chi^2\,T_*^2}\,,
\label{eq:annihilation}
\ee 
where $n_\phi$, the abundance of scalar particles produced from bubble collisions, can be calculated using the formalism described in the previous sections. We will calculate this contribution numerically in the next section, but it is possible to provide some qualitative arguments that the above can at most be an $\mathcal{O}(1)$ correction to DM abundance, as follows: The ratio of the abundances is given approximately by the ratio of the squares of the corresponding couplings, $n_\chi/n_{\phi\,(E_\phi>2m_\chi)}\sim y_\chi^2/\lambda_\phi^2$.   The fraction of high energy $\phi$ states that annihilate into DM states rather than losing their energy through scattering can be roughly estimated to be $\sim y_\chi^2/\lambda_\phi^2$. Therefore, the $\phi$ particles with sufficient energy to produce DM through annihilations are $\sim  \lambda_\phi^2/y_\chi^2$ times as abundant as DM, but only a $\sim y_\chi^2/\lambda_\phi^2$ fraction annihilate into DM, hence this contribution to the DM relic abundance is expected to be an $\mathcal{O}(1)$ effect. Once the $\phi$ particles attain a thermal distribution, or decay into a thermal SM bath, this thermal population does not contain sufficient energy to produce DM. The above estimates hold if all $\phi$ particles participate in annihilations or scatterings; in practice, the number density $n_\phi$ can be sufficiently low that these interactions do not occur frequently as the Universe expands, in which case the DM contribution is correspondingly smaller. 

In addition to the scalar $\phi$, there exists at least one other dark sector particle in the physical spectrum:
either a (pseudo) Goldstone boson (if the broken symmetry is global) or a gauge boson (if the broken symmetry is gauged). Goldstone annihilations are expected to give a contribution comparable to that from the scalar, since annihilations that produce heavy DM occur at energies above the scale of symmetry breaking, where Goldstone interactions are expected to be similar to scalar interactions. Gauge bosons can only annihilate directly to DM provided the DM particle is charged under the symmetry that the gauge boson corresponds to; in this case, the contribution from this process can be calculated in the same way. In the absence of a direct (gauge) coupling, gauge boson annihilation to DM can occur through diagrams mediated by the scalar, but this contribution is expected to be subdominant compared to the abundance produced directly from scalar annihilations. 

In a specific model, the DM abundance from such $\phi$ as well as other dark sector particle annihilations can be obtained by numerically solving the Boltzmann equations with the appropriate injection of the dark sector particle spectra from bubble collisions; however, this is beyond the scope of the present work.

\subsection{Subsequent Evolution}

At production, the DM particles are localized around sites of bubble collisions, but since they are highly boosted, with energies $E_\chi\sim \gamma_w v_\phi$, they quickly propagate over all space and reach a homogeneous distribution. For consistency with observations, the DM population is required to become cold (i.e.\,nonrelativistic) by the time of matter-radiation equality, which should occur when the temperature of the radiation bath drops down to the keV scale. If $E_\chi/m_\chi>T_*/\text{keV}$, the redshift of momenta due to the expansion of the Universe is not sufficient to achieve this, and DM particles need to scatter multiple times with other dark sector particles to become sufficiently cold. To this end, one can check if a DM particle scatters with a $\phi$ particle within a Hubble time after the completion of the phase transition, weighed by the fractional momentum loss from each collision; the condition for this to occur is \cite{Baldes:2022oev}
\be
n_\phi\,\sigma_{\phi\chi\to\phi\chi} v \frac{\delta p_{\text{DM}}}{p_{\text{DM}}}>H ~~~\Rightarrow~~~n_\phi\,\frac{\lambda_\chi^2}{64\, m_\chi^2}\frac{T_n}{T_*}>1.66\sqrt{g_*}\frac{T_*^2}{M_{Pl}}\,,
\ee
where in the second step we have used various relations given in \cite{Baldes:2022oev}.
It can be challenging to satisfy the above condition if DM only has a small coupling $\lambda_\chi$ to the scalar field, or if the $\phi$ number density $n_\phi$ is significantly smaller than a thermal abundance $\sim T^3$.  In such cases, DM cannot dissipate its energy and might be too hot at late times to be consistent with observations. 

If DM does cool sufficiently through a combination of redshift due to the expansion of the Universe and scattering with other dark sector particles, such DM particles can nevertheless feature long free-streaming lengths that can leave observable imprints. Ref.\!\cite{Baldes:2022oev} explored various observational effects of heavy DM produced with large boosts, and found that such long free-streaming lengths for DM can result in a suppressed matter power spectrum that could provide measurable effects for future cosmological observations; see Ref.\!\cite{Baldes:2022oev} for further details. It is also interesting to note that a small fraction of DM, if sufficiently boosted, might also contribute to dark radiation at Big Bang Nucleosynthesis (BBN) (see e.g.~\cite{Shakya:2016oxf}). Finally, it is worth noting that ultraheavy DM close to the Planck scale could potentially also be detected purely through its gravitational interactions with experimental efforts such as the Windchime project \cite{Windchime:2022whs}.

\section{Dark Matter Parameter Space}
\label{sec:parameterspace}

In this section, we explore the parameter space where dark matter can be produced with the desired abundance using the formalism described in the previous sections. We will provide an extensive discussion for the case of scalar DM, and discuss fermion and vector DM, which involve more subtleties (see discussion in Sec.\,\ref{subsec:content}), more briefly. 

First, it is useful to rewrite various relevant expressions and conditions discussed above in terms of the phase transition parameters defined in Sec.\,\ref{parameters}. We assume that the energy released in the phase transition gets converted to a thermal bath of SM and dark sector particles. Eventually all dark sector particles (other than DM) decay into the SM. Using energy conservation, the temperature $T_*$ of this SM bath can be calculated via
\be
\frac{\pi^2}{30}g_* T_*^4=\rho_{\text{radiation}}+\Delta V=\frac{\Delta V}{\alpha}+\Delta V=\frac{1+\alpha}{\alpha}\,c_V\,v_\phi^4\,,
\ee 
where $\rho_{\text{radiation}}$ is the energy density in the radiation bath prior to the phase transition, $T_*$ is the temperature of the thermalized bath, $g_*$ is the number of degrees of freedom (d.o.f.) in the final thermal bath (for which we will use $g_*=100$, which approximates the SM d.o.f.~above the QCD phase transition), and we have used various definitions and relations provided in Sec.\,\ref{parameters}. Thus we have
\be
T_*=\left(\frac{30(1+\alpha)}{g_* \pi^2 \alpha}c_V\right)^{1/4} v_\phi\,.
\ee

For the temperature $T_n$ at which the phase transition commences, we can similarly use 
\be
\frac{\pi^2}{30}g_{*i} T_n^4=\frac{\Delta V}{\alpha}~~~\Rightarrow~~~T_n=\left(\frac{30}{g_{*i} \pi^2 \alpha}c_V\right)^{1/4} v_\phi\,,
\ee
where $g_{*i}$ is now the number of degrees of freedom in the plasma when the phase transition occurs. For simplicity, we will assume that the initial bath is made up of both SM and dark sector particles, and use $g_{*i}\approx g_*= 100$; if the bath only contains the SM or dark sector, this only introduces $\mathcal{O}(1)$ corrections to our final results. Thus we have $T_n/T_*=(1+\alpha)^{-1/4}$. 

We can also rewrite the formula for the DM abundance, Eq.\,\ref{dmgeneral}, in terms of the phase transition parameters from Sec.\,\ref{parameters} and the formula for the number density produced from background field dynamics, Eq.\,\ref{number2}, as
\be
\Omega_\chi h^2\approx5\times 10^{-12} \frac{\beta}{H}\left(\frac{\alpha}{(1+\alpha)g_* c_V}\right)^{1/4}\frac{m_\chi}{v_\phi}\frac{1}{\text{GeV}^2}\,\int_{p^2_{\text{min}}}^{p^2_{\text{max}}} dp^2\,f(p^2) \,\text{Im} [\tilde{\Gamma}^{(2)}(p^2)]\,.
\label{dmbg1}
\ee
As stated earlier, we can use the functional form $f(p^2)=f_{\rm PE}(p^2)$ from Eq.\,\ref{eq:felastic} for both elastic and inelastic collisions, as we assume that DM is heavier than the scalar field $\phi$, hence the second terms in Eqs.\ref{eq:elasticfit},\,\ref{eq:inelasticfit}, corresponding to scalar field oscillations after wall collision, cannot produce DM particles and can be neglected. The logarithmic factor in the expression for $f_{\rm PE}(p^2)$ requires a numerical evaluation of the integral in Eq.\,\ref{dmbg1}, which is cumbersome and time-consuming; a simplification can be made by observing that this logarithmic factor evaluates to a number between $6$ and $60$ across the range of parameter values of interest to us, hence we can reasonably approximate $\text{Log}\left[\frac{2(\gamma_w/l_w)^2-p^2+2(\gamma_w/l_w)\sqrt{(\gamma_w/l_w)^2-p^2}}{p^2}\right]\approx 20$ for all cases to simplify our calculations. This enables us to further simplify the above formula as
\be
\Omega_\chi h^2\approx 0.1\, \left(\frac{\beta/H}{10}\right)\left(\frac{\alpha}{(1+\alpha)g_* c_V}\right)^{1/4}\frac{m_\chi\,v_\phi}{\text{(2.5 TeV)}^2}\,\int_{p^2_{\text{min}}}^{p^2_{\text{max}}} \frac{dp^2}{p^4} \,\text{Im} [\tilde{\Gamma}^{(2)}(p^2)]\,.
\label{dmbg2}
\ee
For a given form of the function $\text{Im} [\tilde{\Gamma}^{(2)}(p^2)]$, it is then possible to solve this integral analytically, thereby obtaining a fully analytic expression for the DM relic abundance. We will do this for various cases in the following subsections. 

\subsection{Gravitational Waves}
\label{sec:gw}

Before delving into the details of DM production, it is worth discussing the connection with gravitational waves. One of the main attractive features of FOPTs in contemporary research is that they can give rise to stochastic GW signals that can be observed with a variety of existing and upcoming GW detectors. It is therefore judicious to examine whether the FOPTs that can produce the correct DM relic abundance can also give sizable GW signals, which would provide a unique observational probe of this DM production mechanism.

FOPTs can produce gravitational waves in several ways: through the scalar field energy densities in the bubble walls after collision\,\cite{Kosowsky:1991ua,Kosowsky:1992rz,Kosowsky:1992vn,Kamionkowski:1993fg,Caprini:2007xq,Huber:2008hg,Bodeker:2009qy,Jinno:2016vai,Jinno:2017fby,Konstandin:2017sat,Cutting:2018tjt,Cutting:2020nla,Lewicki:2020azd}, the production of sound waves\,\cite{Hindmarsh:2013xza,Hindmarsh:2015qta,Hindmarsh:2017gnf,Cutting:2019zws,Hindmarsh:2016lnk,Hindmarsh:2019phv} and turbulence\,\cite{Kamionkowski:1993fg,Caprini:2009yp,Brandenburg:2017neh,Cutting:2019zws,RoperPol:2019wvy,Dahl:2021wyk,Auclair:2022jod} in the surrounding plasma, or through energy transfer to nontrivial spatial configurations of feebly-interacting particles\,\cite{Jinno:2022fom}. In this paper, we are primarily interested in runaway bubble configurations, where the bubble walls carry most of the energy released in the transition, hence the GWs are primarily sourced by bubble wall collisions, i.e.\,the scalar field. For such GWs, we use the peak frequency of the signal today as obtained from the results of \cite{Cutting:2020nla}, which can be expressed as \cite{Baldes:2022oev}
\be
f_\text{peak(GW)}=15~\mu\text{Hz}~\frac{\beta}{H}\,g_*^{1/6}\left(\frac{T_*}{10^3\,\text{GeV}}\right)=20~\mu\text{Hz}~\frac{\beta/H}{g_*^{1/12}}\left(\frac{(1+\alpha)}{ \alpha}c_V\right)^{1/4}\left(\frac{v_\phi}{10^3\,\text{GeV}}\right). 
\ee
Using this relation, we can map the scale of the phase transition $v_\phi$ to the optimal frequencies of various gravitational wave detectors, as shown in Table \ref{tab:GWexp}. The table shows the corresponding scales of FOPTs that provide GW signals that peak at the optimal frequencies of various detectors as determined by the above formula for some reasonable choices of parameters ($\beta/H=10,\,g_*=100,\,\alpha=1,c_V=0.1$). For these parameter choices, we also list the viable window of DM masses that can be produced from bubble collisions for reasonable couplings between DM and the background field (in the range $10^{-4}$ to $1$) in each case, as derived from our calculations below (see Sec.\,\ref{subsec:scalardm}, Fig.\,\ref{fig:scalarrelic}\,; these numbers correspond to the case of scalar DM, but the numbers for fermion or vector DM should be comparable). 

Here, it is worth mentioning that if particle production (including DM production) from bubble collisions is a strong effect, it can affect the subsequent production of GWs, modifying the amplitude as well as shape of the GW signal \cite{Inomata:2024rkt}.

\begin{table}[t]
  \begin{center}
    \begin{tabular}{|c|c|c|c|} 
     \hline
      \textbf{Experiment} & $f_{optimal}$/\textbf{Hz} & $v_\phi$\textbf{/GeV} & $m_{\text{DM}}$/GeV \\
      \hline
      \hline
       Pulsar Timing Arrays (PTAs) \cite{Bringmann:2023opz,Madge:2023dxc} & $10^{-8}$ & 0.1 & $10^{13}-10^{16}$\\
      LISA \cite{LISA:2017pwj} & 0.001 & $10^4$ & $10^{6}-10^{15}$\\
      BBO \cite{Harry:2006fi}, DECIGO \cite{Kawamura:2006up} & $0.1$ & $10^6$ & $10^{5}-10^{13}$\\\
      Einstein Telescope (ET) \cite{Punturo:2010zz}, Cosmic Explorer (CE) \cite{Reitze:2019iox} & 10 & $10^8$ & $10^{6}-10^{10}$\\
      \hline
    \end{tabular}
  \end{center}
   \caption{Peak gravitational wave frequencies, corresponding scales of phase transition, and ranges of viable (scalar) dark matter masses (with couplings in the range $10^{-4}$ to $1$) for various existing and planned gravitational wave experiments.}
    \label{tab:GWexp}
\end{table}

\subsection{Scalar Dark Matter}
\label{subsec:scalardm}

Consider scalar DM $\chi_s$  that couples to the background field $\phi$ via $\frac{1}{4}\lambda_s \phi^2 \chi_s^2$, and can be produced via $\phi^*_p\to\chi_s^2,~ \phi \chi_s^2$. Substituting the expressions from Eqs.\,\ref{2scalar},\,\ref{3scalar} into Eq.\,\ref{dmbg2}, and dropping the phase space factors in these equations to enable the integral to be performed analytically, we derive the following expression for the scalar DM relic abundance
\be
\Omega_\chi h^2\approx 0.1\,\frac{\beta/H}{10}\left(\frac{\alpha}{(1+\alpha)g_* c_V}\right)^{1/4}\frac{\lambda_s^2\,m_{\chi_s}\,v_\phi}{\text{(24 TeV)}^2}\,\left[\frac{v_\phi^2}{m_{\chi_s}^2}+\frac{1}{16\pi^2}\,\text{ln}\left(\frac{2\,\gamma_w/l_{w0}}{(2m_{\chi_s}+m_\phi)}\right)\right]\,.
\label{eq:scalarabundance}
\ee
The two terms in the square parenthesis correspond to contributions from the two- and three-body decays, respectively. We can see that the latter contribution dominates for $m_{\chi_s}\gtrsim 4\pi\, v_\phi$, clearly demonstrating the importance of the three-body decay channel for heavy scalar DM. We have numerically checked that the above analytic result matches the full numerical result (obtained from evaluating Eqs.\,\ref{eq:felastic},\,\ref{2scalar},\,\ref{3scalar} numerically without dropping any factors) up to an $\mathcal{O}(1)$ factor over the parameter space we are interested in. 

\begin{figure}[t]
\centering
\includegraphics[width=0.48\textwidth]{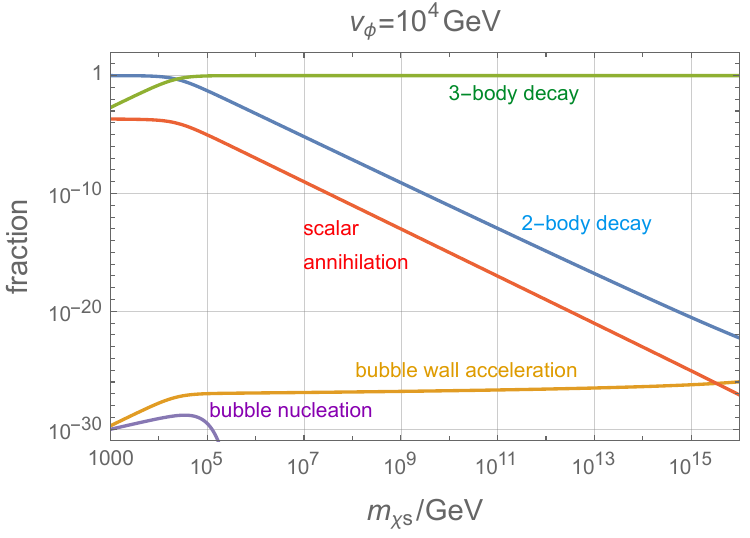}~~
\includegraphics[width=0.48\textwidth]{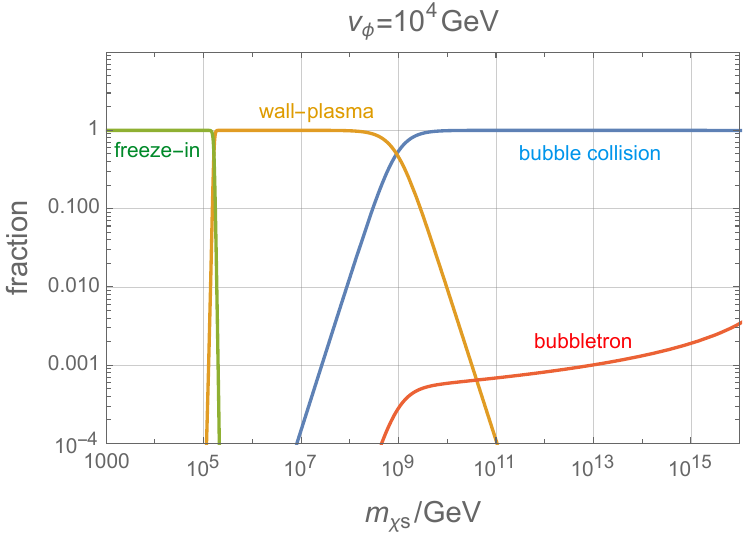}\\
\vskip 0.2cm
\includegraphics[width=0.48\textwidth]{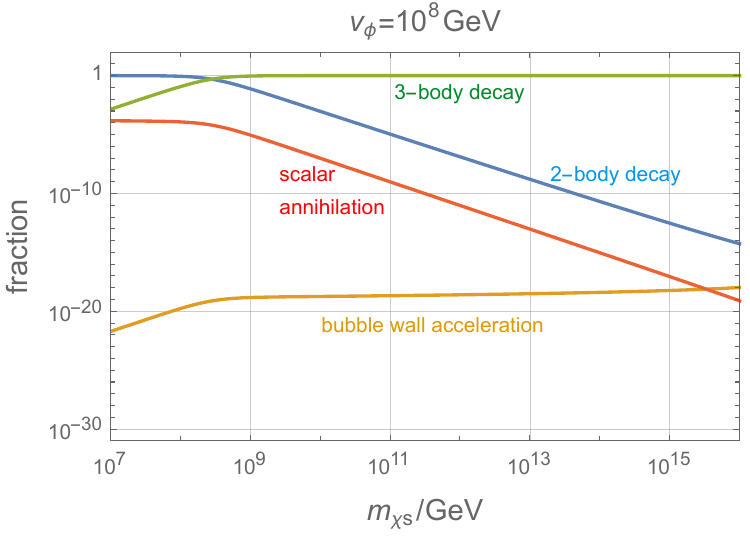}~~
\includegraphics[width=0.48\textwidth]{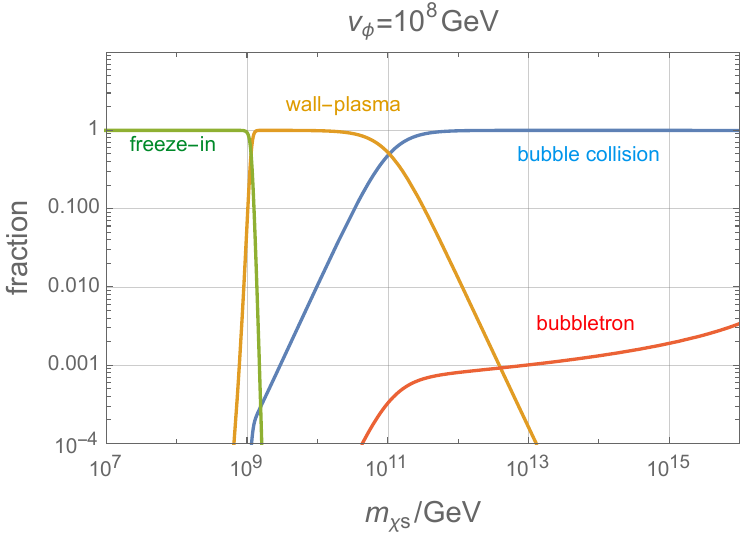}\\
\caption{
Relative contributions to scalar dark matter relic abundance from various processes as a function of dark matter mass, for $v_\phi=10^4$ GeV (top row) and $v_\phi=10^{8}$ GeV (bottom row), in the absence (left panels) or presence (right panels) of a thermal bath. The bubble collision contributions on the right column panels represent the sum of the 2- and 3-body decay contributions from the corresponding left column panels. See text for further details. 
}
\label{fig:scalarcontributions}
\end{figure}

As discussed in the previous sections, several processes contribute to DM production in FOPTs in addition to the background field dynamics at bubble collision: bubble nucleation, bubble expansion (bubble wall acceleration), and annihilations of dark sector particles produced from bubble collisions in all cases (with or without a thermal bath), as well as freeze-in from the thermal bath, wall-plasma interactions, and collisions of accelerated particle shells in the presence of a thermal bath of particles. In Fig.\,\ref{fig:scalarcontributions}, we plot the relative weights of these contributions in the final DM relic density as a function of DM mass for various parameters choices in the absence (left column) or presence (right column) of a thermal bath of particles. We have chosen $v_\phi=10^4$ GeV in the top panel (the relevant scale for a GW signal observable by LISA) and $v_\phi=10^8$ GeV in the bottom panel (the appropriate scale for Einstein Telescope / Cosmic Explorer). In all cases, the coupling $\lambda_s$ has been chosen such that the sum of all contributions produces the correct DM relic density. For these plots, we have chosen the following values for the various parameters:
\be
c_V = 0.1,~\beta/H=10,~\alpha=1,~ R_0=10/v_\phi,~ l_{w0}=1/v_\phi\,.
\label{paramchoice}
\ee
We have assumed that the bubble walls are in the runaway regime throughout the bubble expansion phase, and with the  parameters of Eq.\,\ref{paramchoice} the wall boost factor at the time of collision is
\be
\gamma_w=\frac{2 R_*}{3\,R_0}\approx  \frac{0.15}{\beta/H}\frac{M_{Pl}}{v_\phi}\,,
\label{gammachoice}
\ee
where we have used the relations and parameters listed above. Recall that in the presence of a light gauge boson, the boost factor can reach a terminal value smaller than the above expression (see Eq.\,\ref{gammalightg}); however, we do not consider this possibility in this paper.

In the left panels, we plot the two- and three-body decay contributions from $\phi^*\to\chi^2,~ \phi \chi^2$ separately, in blue and green respectively. It can be seen that the former dominates for $m_\chi\lesssim4\pi v_\phi$, whereas the latter takes prominence for higher DM masses, as anticipated from earlier discussions. The contribution from the annihilation of $\phi$ particles produced from bubble collisions  into DM (assuming the scalar quartic coupling $\lambda_\phi=1$), corresponding to Eq.\,\ref{eq:annihilation}, is shown in red; this contribution is always found to be a few orders of magnitude smaller than the two-body contribution, and therefore subdominant. Likewise, we also plot the contributions from the bubble nucleation and bubble wall acceleration phases (Sec.\,\ref{sec:formalism}), in purple and gold respectively. These contributions are seen to be several orders of magnitude smaller than that from bubble collision due to the reasons discussed in Sec.\,\ref{sec:formalism} and therefore completely negligible. These patterns continue to hold as $v_\phi$ is increased from $10^4$ to $10^8$ GeV (top to bottom left panel). The wall acceleration contribution increases by a few orders of magnitude since it scales as $R_*^{-3}\sim H^3$ and the Hubble scale is higher for higher $v_\phi$, but remains negligible. Meanwhile, the bubble nucleation contribution gets further suppressed as it does not rise as rapidly as the other contributions with $v_\phi$. Therefore, in the absence of a thermal bath, we find that bubble collisions are the dominant DM production mechanism for any choice of parameters.  

In the right panels, we show the relative contributions from various processes in the presence of a thermal bath. We now combine the two- and three-body decays into a single contribution, denoted by the blue curves labelled ``bubble collision". As discussed in the previous section, the presence of a thermal bath introduces several new DM production mechanisms: freeze-in from the annihilation of the scalar particles present in the thermal bath $\phi\phi\to\chi_s\chi_s$ (Eq.\,\ref{renormalizableIR}), denoted by the green curves labelled ``freeze-in"; non-adiabatic transition of $\phi$ to $\chi_s\chi_s$ when $\phi$ particles from the plasma interact with the bubble walls (Eq.\,\ref{eq:nonadiabatic}), denoted by the gold curves labelled ``wall-plasma"; and the collisions of boosted $\phi$ particle shells (Eq,\,\ref{eq:bubbletron}), denoted by the red curves labelled ``bubbletron". We see that the freeze-in contribution from annihilations of $\phi$ particles in the thermal bath can dominate for $m_{\chi_s}\lesssim v_\phi$, but rapidly becomes ineffective for $m_{\chi_s}\gtrsim v_\phi$ as the exponential suppression in (Eq.\,\ref{renormalizableIR}) becomes significant (here we chose $T_R\approx v_\phi$; for larger $T_R$, the freeze-in contribution is expected to dominate for $m_{\chi_s}\lesssim T_R$). The non-adiabatic wall-plasma interaction contribution dominates in the intermediate DM mass regime, as long as
\be
\frac{1}{\beta/H}\left(\frac{m_{\chi_s}/v_\phi}{10^5}\right)^2\frac{v_\phi}{10^4 ~\text{GeV}} \lesssim 1\,.
\ee 
Raising $m_{\chi_s}/v_\phi$ results in the suppression of the effective mixing angle between $\phi$ and $\chi_s\chi_s$ (see discussion surrounding Eq.\,\ref{nonadiabatic}), which eventually makes this contribution subdominant to the abundance produced from three-body decays of the background field excitations. For higher $m_{\chi_s}$ values, bubble collision therefore becomes the dominant DM production process. The bubbletron contribution is seen to grow in importance with increasing $m_{\chi_s}$ but does not dominate in any part of the parameter space, only contributing at percent level at best; however, annihilation cross-sections at high energies scale as $\sim E^{-2}\sim(\gamma_w v_\phi)^{-2}$ (using $\gamma_w$ from Eq.\,\ref{gammachoice}), so it is possible that a configuration with smaller $\gamma_w$ could enhance the bubbletron contribution and make it relevant in some parts of parameter space. Note that the above contributions exist in the presence of a plasma, but can be suppressed in a supercooled phase transition, where a brief inflationary phase due to vacuum domination can lead to significant dilution of the pre-existing thermal bath, thereby suppressing these contributions. 

Here it is worth emphasizing that previous studies only considered the two-body decay channel $\phi^*_p\to\chi_s\chi_s$ and hence underestimated particle production from bubble collisions, finding it to be subdominant to the wall-plasma or bubbletron contribtutions. When the three-body decay process $\phi^*_p\to\phi\chi_s\chi_s$ is taken into account, we find that this can constitute the dominant contribution for heavy DM even in the presence of a thermal bath and the above processes. 
 
\begin{figure}[t]
\centering
\includegraphics[width=0.7\textwidth]{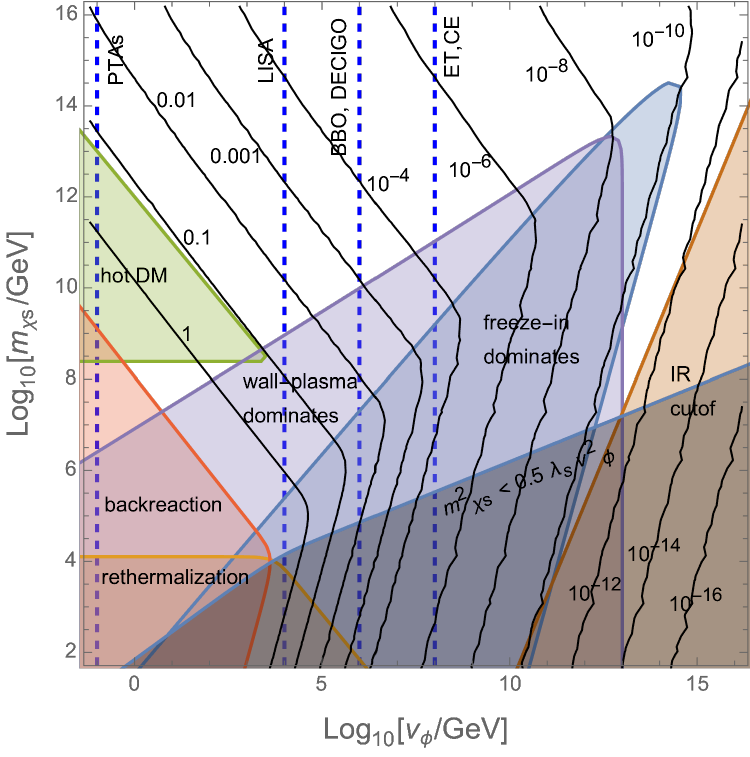}
\caption{
Contours of the value of the coupling $\lambda_s$ for which the correct scalar dark matter relic density is achieved, as a function of the dark matter mass and the scale of symmetry breaking. The vertical dashed blue lines denote the transition scales that produce gravitational waves at peak sensitivity frequencies for various GW detectors (see Sec.\,\ref{sec:gw}). The shaded regions represent various constraints, see text for detailed discussions. Note that the blue (``freeze-in dominates") and purple (``wall-plasma dominates") regions are only applicable if a thermal dark sector bath is present before the phase transition, whereas the other regions are applicable for all FOPTs, with or without a bath. 
}
\label{fig:scalarrelic}
\end{figure}

Having examined the relative importance of various processes, we next plot, in Fig.\,\ref{fig:scalarrelic}, contours of the magnitude of the coupling $\lambda_s$ required to produce the correct DM relic abundance from bubble collisions as a function of $v_\phi$ and $m_{\chi_s}$, together with various constraints. This figure shows that bubble collisions can account for the correct DM relic density across a vast range of scales spanning several orders of magnitude. In this plot, we have restricted $v_\phi$ to values above 100 MeV (below this, an FOPT is likely to disrupt BBN) and $m_{\chi_s}$ to values above 100 GeV (lower values are possible, but there are no qualitatively new features beyond what is already seen in the figure). On both axes, we implement an upper cutoff of $10^{16}$ GeV; beyond this, the value approaches the UV-cutoff of the system (given by the bubble wall thickness at collision $\gamma_w/l_{w 0}$), and details of the bubble wall profile, which we have not taken into account, become important. The dashed vertical lines denote the FOPT symmetry breaking scales corresponding to the peak sensitivities of various gravitational wave experiments, as described in Sec.\,\ref{sec:gw} (see Table \ref{tab:GWexp}). We see that ultraheavy DM several orders of magnitude heavier than the symmetry breaking scale $v_\phi$ can be produced for reasonable values of the coupling $\lambda_s$ across a large range of FOPT scales of interest for various current and planned GW experiments. It should be emphasized that the DM mass could also be significantly larger than the highest temperature ever reached in our cosmic history. 

In the figure, we see that all the contours feature sharp kinks at $m_{\chi_s}\sim 4\pi v_\phi$. This corresponds to the transition between regions of parameter space where two- or three-body decays of the background field excitations dominate the DM production process. For $m_{\chi_s}\lesssim 4\pi v_\phi$, the two-body process dominates;  from Eq.\,\ref{eq:scalarabundance}, $\Omega_\chi h^2\propto 1/m_{\chi_s}$ when the first term in the parenthesis dominates, hence as $m_{\chi_s}$ is increased for fixed $v_\phi$, the required coupling increases. In the opposite regime $m_{\chi_s}\gtrsim 4\pi v_\phi$, the three-body process dominates, and the relic density scaling from Eq.\,\ref{eq:scalarabundance} when the second term in the parenthesis dominates is instead $\Omega_\chi h^2\propto m_{\chi_s}$, and now the required coupling decreases as $m_{\chi_s}$ is increased for fixed $v_\phi$. This explains the reversal of the contour shapes across the two regions in the plot. 

The shaded regions denote various constraints, as follows:

\begin{itemize}

\item The red region in the bottom left denotes parameter space where the backreaction from particle production becomes important, given approximately by the condition $\lambda_s>10$; note that this also corresponds to a regime where the coupling becomes nonperturbative. 

\item The green triangular region denotes parameter space where DM is too hot to account for the structures observed today. The constraint disappears for higher $m_{\chi_s}$ because heavier DM is not as boosted and can redshift to become cold by the time of matter-radiation equality. Similarly, the constraint also disappears below $m_{\chi_s}\approx10^{8.5}$ GeV because the value of $\lambda_s$ required to produce the correct relic density increases, enabling more efficient $\chi_s -\phi$ scattering, which disperses the energy carried by DM particles, allowing them to cool. 

\item The orange region in the bottom left represents parameter space where the produced DM population re-enters chemical equilibrium with the dark sector bath (i.e. the rate for $\chi_s\chi_s\to \phi\phi$ scattering is faster than Hubble), in which case imprints of the bubble collision process are washed out and DM subsequently undergoes thermal freezeout. Note that this region disappears to the right even though $m_{\chi_s}\ll v_\phi$ since the associated coupling $\lambda_s$ also becomes very small. 

\item The brown region in the bottom right corresponds to configurations where $m_{\chi_s}< R_*^{-1}$, i.e. the formalism used to calculate particle production from bubble collisions falls below its IR cutoff and is no longer valid, since the existence of multiple bubbles should be taken into account in the Fourier transform of the background field.

 \item The blue region corresponds to parameter space where freeze-in from the annihilation of the scalar particles $\phi\phi\to\chi_s\chi_s$ (Eq.\,\ref{renormalizableIR}) from the pre-existing thermal bath, if present, dominates over production from bubble collision. Note that its upper boundary occurs at roughly $m_{\chi_s}\sim v_\phi$; for heavier DM masses, the exponential suppression in Eq.\,\ref{renormalizableIR} rapidly suppresses this contribution (recall that we chose $T_R\sim v_\phi$; for larger $T_R$ this region is expected to get bigger). Likewise, the lower boundary of this region coincides approximately with the $\lambda_s\sim 10^{-10}$ contour; as we can see from Eq.\,\ref{renormalizableIR}, even in the absence of the exponential, freeze-in cannot provide the correct relic density for smaller values of $\lambda_s$. 
 
\item  The purple region represents parameter space where the non-adiabatic transition of $\phi$ to $\chi_s\chi_s$ due to $\phi$ particles from the plasma (if it exists) interacting with the bubble walls (Eq.\,\ref{eq:nonadiabatic}) dominates over the bubble collision contribution. Beyond the upper boundary and to the right of this region, the non-adiabaticity condition from Eq.\,\ref{nonadiabatic} continues to hold, but production becomes less efficient than that from bubble collisions. Where blue and purple regions overlap, the blue (freeze-in) contribution generally tends to be larger (see Fig.\,\ref{fig:scalarcontributions}).

\item As discussed earlier, we grey out the region $m_{\chi_s}^2\!<\!\frac{1}{2}\lambda_s v_\phi^2$, where the mass contribution expected from the phase transition exceeds the DM mass. In this part of the parameter space, higher order insertions beyond the leading contribution considered in this paper can also become important, as discussed in Sec.\,\ref{others}. Furthermore, in this region DM is likely lighter than $\phi$ as well as other dark sector states, hence the assumption that decays of dark sector particles do not produce DM also possibly breaks down. In addition, it should be kept in mind that we have used $f(p^2)=f_{\rm PE}$ for our calculations, which assumes $m_\chi > m_\phi$; in regions of parameter space where this is not the case, additional contributions from the scalar oscillations after bubble collisions would likely dominate and overproduce DM.   

 \end{itemize}
 
Note that Eq.\,\ref{eq:scalarabundance} can be further simplified by approximating the second parenthesis as $1$ when $\alpha\geq 1$ and the log factor as $10$ over the region of parameter space shown in Fig.\,\ref{fig:scalarrelic}. In the regime where three-body decays dominate (generally the case for $m_{\chi_s}\gtrsim 4\pi v_\phi$), this yields the following simple relation among the parameters to achieve the correct DM relic density
\be
\frac{\lambda_s^2\,\beta/H}{10}\frac{m_{\chi_s}\,v_\phi}{\text{(100 TeV)}^2}\approx 1.
\label{eq:scalarabundancesimplified}
\ee
This is found to be in excellent agreement with the results shown in Fig.\,\ref{fig:scalarrelic}. 

Finally, we provide an intuitive discussion of  the number density of heavy DM particles produced from bubble collisions, based on the results derived above (in particular Eqs.\,\ref{number2},\,\ref{dmgeneral},\,\ref{dmbg1},\,\ref{eq:scalarabundance}). Parametrically, we see that the number of particles produced per unit area of colliding bubble walls scales as $\sim \lambda_s^2\, v_\phi^2$ (times a logarithmic factor), with no other parametric suppression. This diffuses over the size of the bubble to yield a number density  $n\sim \lambda_s^2\,v_\phi^2/R_*$; since $l_{w0}\sim v_\phi^{-1}$ and $T_*\sim v_\phi$, this can be rewritten as $n\sim \lambda_s^2\, T_*^3 (l_{w0}/R_*)$. This invites the interpretation that for $\mathcal{O}(1)$ couplings, bubble collisions are efficient at producing essentially a thermal abundance $\sim T_*^3$ of particles within the extent of the bubble wall $l_{w0}$ (as seen in its rest frame), which then diffuses out throughout the bubble. One can also understand the above parametric scaling through a different argument: the energy released in the phase transition is $\Delta V\sim v_\phi^4$, whereas bubble collisions produce particles with typical energy $\gamma_w v_\phi$. Hence efficient particle production from bubble collisions should lead to a number density $n\sim v_\phi^3/\gamma_w$. Recalling that $\gamma_w\sim R_*/R_0\sim R_*/l_{w0}$, this is the same number density as above for $\mathcal{O}(1)$ couplings.

\subsection{Fermion Dark Matter}
\label{sec:fermiondm}

We now consider the case of fermion DM. Many of the qualitative details and discussions are similar to the case of scalar DM, and we will not repeat them here but instead focus on the differences. 

The fermion DM relic abundance from $\phi^*_p\to \chi_f\bar{\chi}_f$ decays arising from the interaction term $y_f\phi\chi_f\bar{\chi}_f$ can be expressed as
\be
\Omega_\chi h^2\approx0.1\, \frac{\beta/H}{10}\left(\frac{\alpha}{(1+\alpha)g_* c_V}\right)^{1/4}\frac{m_{\chi_f}\,v_\phi}{\text{(8.6 TeV)}^2}\,y^2_f\,\text{ln}\left(\frac{\gamma_w/l_{w0}}{m_{\chi_f}}\right)\,.
\label{eq:fermionabundance}
\ee

\begin{figure}[t]
\centering
\includegraphics[width=0.7\textwidth]{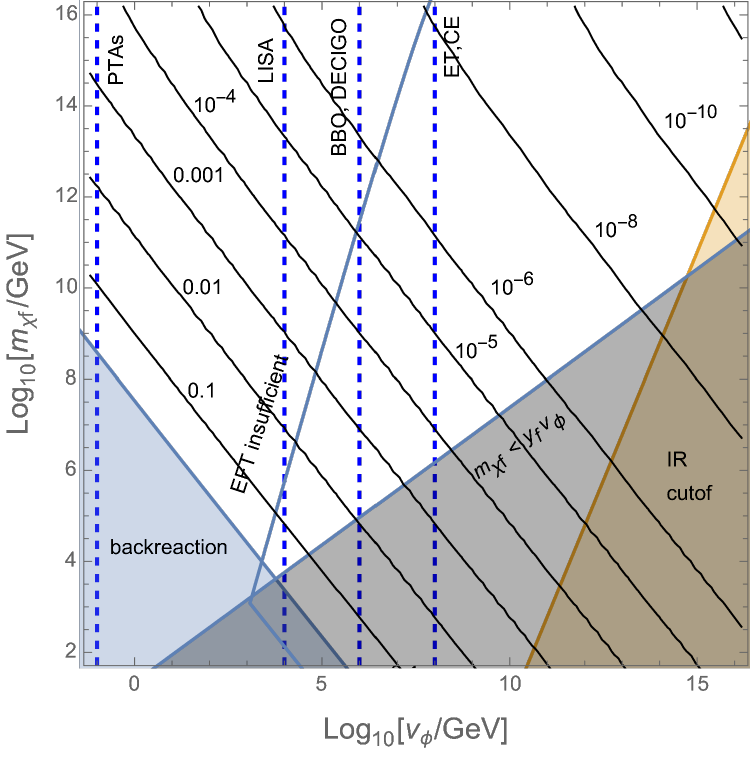}
\caption{
Contours of the value of the coupling $y_f$ for which the correct fermion DM relic density is achieved, as a function of the scale of symmetry breaking and the DM mass. The vertical dashed blue lines denote the transition scales that produce gravitational waves at peak sensitivity frequencies for various detectors (see Sec.\,\ref{sec:gw}). Details of the constraints in the various regions are discussed in the text. 
}
\label{fig:fermionrelic}
\end{figure}

In Fig.\,\ref{fig:fermionrelic}, we plot contours of the magnitude of the coupling $y_f$ required to produce the correct fermion DM relic abundance as a function of $v_\phi$ and $m_{\chi_f}$, analogous to the scalar DM case in Fig.\,\ref{fig:scalarrelic}. We choose the same parameters as for scalar DM (Eq.\,\ref{paramchoice}), with the boost factor given by Eq.\,\ref{gammachoice}. As in Fig.\,\ref{fig:scalarrelic}, we denote the exclusion region (shaded golden) where the DM mass falls beyond the IR cutoff and the Fourier transform of a single bubble collision is no longer sufficient, and the region (shaded blue) where backreaction from particle production becomes important, i.e.~the energy density in the DM particles (calculated using Eq.\,\ref{energy}) exceeds the latent energy released in the phase transition, $\Delta V$. The latter region is found to correspond to $y_f> 1.5$. Note that this backreaction region for fermion DM is larger than the corresponding region for the scalar DM case in Fig.\,\ref{fig:scalarrelic} (corresponding approximately to $\lambda_s>10$) since the fermion is produced via the two-body decay channel $\phi^*_p\to\chi_f\bar{\chi}_f$, whereas the leading contribution for scalar DM at high energies is the phase-space suppressed three-body decay $\phi^*_p\to\phi\chi_s^2$. Also note that the contours of constant $y_f$ in Fig.\,\ref{fig:fermionrelic} are straight lines, as two-body decay is the only relevant process for fermion DM, and do not feature the kinks observed in Fig.\,\ref{fig:scalarrelic} due to two- and three- body decays being important in different parts of parameter space for scalar DM. Similarly, we show the boundary $m_{\chi_f} = y_f v_\phi$ below which higher order corrections become important and need to be included to get the correct result (see Sec.\,\ref{others}). As with the scalar DM case, we emphasize that we have used $f(p^2)=f_{\rm PE}$ for our calculations, which assumes $m_\chi > m_\phi$; in regions of parameter space with $m_\chi < m_\phi$, production from oscillations of the scalar field, not considered in the above calculations, will dominate the DM abundance. 

Unlike Fig.\,\ref{fig:scalarrelic}, here we do not plot any constraints arising from interactions, such as hot DM, rethermalization, or freeze-in constraints, since these involve $2\to 2$ interactions, whose nature depends on the UV-completion of the $y_f\phi\chi_f\bar{\chi}_f$ term. As discussed in Sec.\,\ref{subsec:content}, if $\chi_f$ is not charged under the symmetry broken by $\phi$, the above interaction is an effective field theory (EFT) interaction derived from some higher dimensional operator of the form $\frac{1}{\Lambda_f}\phi^2\chi_f\bar{\chi}_f$, obtained by integrating out some UV physics at the scale $\Lambda_f$ to give the low energy effective coupling $y_f=v_\phi/\Lambda_f$. In the plot, we denote the region (labelled ``EFT insufficient") where $\Lambda_f< \text{min}(m_{\chi_f},\,v_\phi)$, which corresponds to the breakdown of this EFT, where the new degrees of freedom are no longer heavy and cannot be integrated out. This does not mean that fermion DM with such masses cannot be produced with FOPTs at these scales, but simply that additional physics beyond the minimal EFT interaction considered above would be relevant and could provide the leading effect, hence the EFT calculation can no longer be trusted to give the correct result. Note that these new d.o.f. at the scale $\Lambda_f$ also affect the result in the remainder of parameter space where $\Lambda_f> m_{\chi_f},v_\phi$: the particle production calculation involves evaluating the integral in Eq.\,\ref{dmbg2}, which runs up to $p_{\text{max}}\approx E_w$. Since the EFT interaction term is only valid up to $p=\Lambda_f$, the new d.o.f. would modify the integrand at larger $p$ values; however, since the integral evaluates to a logarithm (see Eq.\,\ref{eq:fermionabundance}), we expect this to modify the result only by an $\mathcal{O}(1)$ factor. 

Finally, we can also obtain a very simple approximate relation among the parameters that achieves the correct fermion DM relic density, analogous to Eq.\,\ref{eq:scalarabundancesimplified} and obtained by making the same approximations:
\be
 \frac{y^2_f\,\beta/H}{10}\frac{\,m_{\chi_f}\,v_\phi}{\text{(2.7 TeV)}^2}\approx 1.
\label{eq:fermionabundancesimp}
\ee

\subsection{Vector Dark Matter}
\label{sec:vector}

Next, we briefly discuss the case of a vector DM particle $\chi_v$, which gets produced via the effective operator $\frac{1}{2}\lambda_V\phi^2 \chi_v^\mu \chi_{v\,\mu}$ introduced in Sec.\,\ref{subsec:content}.  The production of vector DM is similar to the case of scalar DM discussed in Sec.\,\ref{subsec:scalardm} and Fig.\,\ref{fig:scalarrelic} (since it gets produced through two- and three-body decays $\phi^*_p\to\chi_v\chi_v,\, \phi\chi_v\chi_v$ similar to the scalar case), with the same UV-completion caveats as for fermion DM (Sec.\,\ref{sec:fermiondm}). We therefore do not make a separate plot for vector DM or repeat similar discussions, but simply mention the additional aspects relevant for the vector case.

If $\chi_v$ is the gauge boson corresponding to the symmetry broken by the $\phi$ vev (in which case $m_{\chi_v}=g\,v_\phi\lesssim v_\phi$, , where $g$ is the gauge coupling), the details of production of transverse and longitudinal modes map directly to the discussion in Sec.\,\ref{sec:gaugedependence}, with $\lambda_V\to g^2$. In this case, for the two-body decay $\phi^*_p\to\chi_v\chi_v$ one can interpolate between the low energy ($p^2<m_{\chi_v}^2$) behavior given by Eq.\,\ref{MgaugeU} and the high energy ($p^2>m_{\chi_v}^2$) behavior obtained from the Goldstone Equivalence Theorem, Eq.\,\ref{GETresult}, to construct an approximate solution valid across all energy scales. For the three-body decay $\phi^*_p\to\phi\chi_v\chi_v$, one can similarly interpolate between the two expressions in Eq.\,\ref{3sg} to construct an approximate solution across all energies. Note that the production rate now depends not only on the gauge coupling $g$ but also the scalar quartic coupling $\lambda_\phi$, as the longitudinal component of the gauge boson behaves as the Goldstone field at high energies, as anticipated from the GET.  

For $m_{\chi_v}\gg g v_\phi$, $\chi_v$ cannot be the gauge boson of the broken symmetry and hence does not couple directly to $\phi$, and the effective operator $\frac{1}{2}\lambda_V \phi^2 \chi_v^\mu \chi_{v\,\mu}$ must arise from integrating out some mediator fields at some scale $\Lambda_v$. As discussed for the case of fermion DM, the new d.o.f.~can modify the results obtained from the EFT operator, and even invalidate the EFT approach in some regions of parameter space. For vector DM, this also introduces an additional consideration: as discussed in Sec.\,\ref{sec:gaugedependence}, a practical approach to obtain physical, pathology-free results for the production of vectors, which might otherwise be gauge-dependent, is to use the Goldstone Equivalence Theorem to obtain the high energy behavior of longitudinal mode production. This requires appropriately replacing the longitudinal mode with the corresponding Goldstone mode at high energies in the UV theory. 

Overall, we expect the production of vector DM to be as general and efficient as the production of scalar DM (from Fig.\,\ref{fig:scalarrelic}), but any model-specific study of vector DM production from bubble collisions must properly address the various aspects discussed above.

\section{Summary and Discussions}
\label{sec:summary}

Here, we summarize the main points of our paper: 

\begin{itemize}
\item We have studied nonthermal production of ultraheavy dark matter (DM) from background field dynamics during a first-order phase transition (FOPT), dominated by bubble collisions, where the bubble walls achieve runaway behavior. This constitutes an unavoidable contribution to DM abundance that exists in any FOPT, irrespective of the nature of the transition or the plasma, when DM couples directly or indirectly to the background field undergoing the transition. The contribution studied in this paper can constitute the dominant production mechanism for heavy DM even in the presence of a thermal plasma and the existence of other well-known DM production mechanisms. 

\item This mechanism is very general and can produce the correct relic density of scalar, fermion, or vector DM across a large range of masses,  from $\mathcal{O}(10)$ TeV to a few orders of magnitude below the Planck scale. This broad regime of validity is the result of the DM number density from bubble collisions being only logarithmically sensitive to the DM mass, resulting in a milder DM mass dependence for the DM relic density (see e.g.~Eq.\,\ref{eq:scalarabundance},~Eq.\,\ref{eq:fermionabundance}) compared to other traditional production mechanisms. 

\item Such setups provide a natural configuration to produce ultraheavy DM with masses many orders of magnitude greater than the scale of the phase transition as well as the temperature of the thermal bath at any point in cosmological history -- for instance, a phase transition at the GeV scale can produce DM as heavy as $\sim 10^{16}$ GeV (see e.g.\,Fig.\,\ref{fig:scalarrelic}).

\item We have demonstrated that the existing formalism for the calculation of particle production in a gauge theory is not gauge invariant, and can lead to spurious results if not treated carefully (see Sec.\,\ref{sec:gaugedependence}). We offer a practical prescription to avoid these complications that makes use of explicit polarization vectors and the Goldstone Equivalence Theorem to extricate physically relevant contributions while avoiding spurious unphysical components. 

\item We have pointed out the importance of three-body decays of the background field excitations, generally ignored in the literature, for the production of scalar and vector particles, which dominate over two-body decays at large energies and provide the dominant contributions for heavy scalar and vector DM. Such three-body decays can also be relevant for fermion DM if the interaction between the background field and fermion DM is mediated by scalar or vector states. 

\item Although DM from bubble collisions does not require any significant couplings to Standard Model particles, which hampers the detection prospects from traditional direct and indirect DM searches, three phenomenological aspects are noteworthy. (i)  FOPTs that produce the correct DM relic abundance can also produce observable gravitational wave signals across a large range of frequencies relevant for current and planned gravitational wave experiments (see Fig.\,\ref{fig:scalarrelic}, Fig.\,\ref{fig:fermionrelic}). (ii) DM produced from bubble collisions is highly boosted, which gives rise to a modified matter power spectrum that could be detectable with future cosmological observations. (iii) Ultraheavy DM with mass close to the Planck scale, which can readily be produced via the mechanism discussed in this paper, could be detectable purely through its gravitational couplings.  

\end{itemize}

We conclude by discussing the broader implications of our work and highlighting several specific directions that could benefit from further study. We have provided model-independent results for heavy DM production from a FOPT with runaway bubbles, which was found to be viable over a large region of parameter space spanning several orders of magnitude. These results have broad applicability and can be implemented in a straightforward manner in specific models of FOPTs, hence it will be interesting to check whether various well-motivated BSM scenarios that give rise to FOPTs with runaway bubbles can be extended to include heavy DM candidates. At the same time, the gauge dependence of the formalism suggests that it is not on rigorous footing, and awaits additional theoretical developments towards a proper completion. It would also be interesting to study the effects of particle production from bubble collisions on the subsequent generation of gravitational wave signals. Various cosmological and astrophysical aspects of ultraheavy boosted DM from bubble collisions, in particular the observational aspects of a modified matter power spectrum, are also worth exploring in greater detail. Beyond DM, the formalism for particle production from bubble collisions developed here could also find applications in other open questions in particle physics, such as baryogenesis or leptogenesis \cite{Cataldi:2024pgt}. We leave the pursuit of such questions for future work. 

\section*{Acknowledgments}

We are grateful to Thomas Konstandin, Henda Mansour, Geraldine Servant, Ville Vaskonen, and Andrea Wulzer for insightful discussions. The work of H.M.L is supported in part by Basic Science Research Program through the National Research Foundation of Korea (NRF) funded by the Ministry of Education, Science and Technology
(NRF-2022R1A2C2003567 and NRF-2021R1A4A2001897). A.P. has been partly supported
by the Departament de Recerca i Universitats from Generalitat de Catalunya to the Grup de Recerca ”Grup de F\'isica Te\`orica UAB/IFAE” (Codi: 2021 SGR 00649) and the research grant PID2020-115845GB-I00/AEI/10.13039/501100011033.
 B.S. is supported by the Deutsche Forschungsgemeinschaft under Germany’s Excellence Strategy - EXC 2121 Quantum Universe
- 390833306. B.S. also thanks the CERN Theory Department and the Lawrence Berkeley
National Laboratory for hospitality during multiple visits over the duration of completion of this project.

\bibliography{DarkMatterPhaseTransition}{}

\providecommand{\href}[2]{#2}\begingroup\raggedright\begin{thebibliography}{100}

\bibitem{Hogan:1983ixn}
C.~J. Hogan, ``{NUCLEATION OF COSMOLOGICAL PHASE TRANSITIONS},''
  \href{http://dx.doi.org/10.1016/0370-2693(83)90553-1}{{\em Phys. Lett. B}
  {\bfseries 133} (1983) 172--176}.

\bibitem{Witten:1984rs}
E.~Witten, ``{Cosmic Separation of Phases},''
  \href{http://dx.doi.org/10.1103/PhysRevD.30.272}{{\em Phys. Rev. D}
  {\bfseries 30} (1984) 272--285}.

\bibitem{Hogan:1986qda}
C.~J. Hogan, ``{Gravitational radiation from cosmological phase transitions},''
  {\em Mon. Not. Roy. Astron. Soc.} {\bfseries 218} (1986) 629--636.

\bibitem{Kosowsky:1991ua}
A.~Kosowsky, M.~S. Turner, and R.~Watkins, ``{Gravitational radiation from
  colliding vacuum bubbles},''
  \href{http://dx.doi.org/10.1103/PhysRevD.45.4514}{{\em Phys. Rev. D}
  {\bfseries 45} (1992) 4514--4535}.

\bibitem{Kosowsky:1992rz}
A.~Kosowsky, M.~S. Turner, and R.~Watkins, ``{Gravitational waves from first
  order cosmological phase transitions},''
  \href{http://dx.doi.org/10.1103/PhysRevLett.69.2026}{{\em Phys. Rev. Lett.}
  {\bfseries 69} (1992) 2026--2029}.

\bibitem{Kosowsky:1992vn}
A.~Kosowsky and M.~S. Turner, ``{Gravitational radiation from colliding vacuum
  bubbles: envelope approximation to many bubble collisions},''
  \href{http://dx.doi.org/10.1103/PhysRevD.47.4372}{{\em Phys. Rev. D}
  {\bfseries 47} (1993) 4372--4391},
  \href{http://arxiv.org/abs/astro-ph/9211004}{{\ttfamily
  arXiv:astro-ph/9211004}}.

\bibitem{Kamionkowski:1993fg}
M.~Kamionkowski, A.~Kosowsky, and M.~S. Turner, ``{Gravitational radiation from
  first order phase transitions},''
  \href{http://dx.doi.org/10.1103/PhysRevD.49.2837}{{\em Phys. Rev. D}
  {\bfseries 49} (1994) 2837--2851},
  \href{http://arxiv.org/abs/astro-ph/9310044}{{\ttfamily
  arXiv:astro-ph/9310044}}.

\bibitem{Guth1981}
A.~H. Guth, ``Inflationary universe: A possible solution to the horizon and
  flatness problems,'' \href{http://dx.doi.org/10.1103/PhysRevD.23.347}{{\em
  Phys. Rev. D} {\bfseries 23} (Jan, 1981) 347--356}.
  \url{https://link.aps.org/doi/10.1103/PhysRevD.23.347}.

\bibitem{LaSteinhardt1989}
D.~La and P.~J. Steinhardt, ``Extended inflationary cosmology,''
  \href{http://dx.doi.org/10.1103/PhysRevLett.62.376}{{\em Phys. Rev. Lett.}
  {\bfseries 62} (Jan, 1989) 376--378}.
  \url{https://link.aps.org/doi/10.1103/PhysRevLett.62.376}.

\bibitem{Grojean:2006bp}
C.~Grojean and G.~Servant, ``{Gravitational Waves from Phase Transitions at the
  Electroweak Scale and Beyond},''
  \href{http://dx.doi.org/10.1103/PhysRevD.75.043507}{{\em Phys. Rev.}
  {\bfseries D75} (2007) 043507},
\href{http://arxiv.org/abs/hep-ph/0607107}{{\ttfamily arXiv:hep-ph/0607107
  [hep-ph]}}.
%%CITATION = HEP-PH/0607107;%%.

\bibitem{Caprini:2015zlo}
C.~Caprini {\em et~al.}, ``{Science with the space-based interferometer eLISA.
  II: Gravitational waves from cosmological phase transitions},''
  \href{http://dx.doi.org/10.1088/1475-7516/2016/04/001}{{\em JCAP} {\bfseries
  1604} (2016) 001},
\href{http://arxiv.org/abs/1512.06239}{{\ttfamily arXiv:1512.06239
  [astro-ph.CO]}}.
%%CITATION = ARXIV:1512.06239;%%.

\bibitem{Caprini:2018mtu}
C.~Caprini and D.~G. Figueroa, ``{Cosmological Backgrounds of Gravitational
  Waves},'' \href{http://dx.doi.org/10.1088/1361-6382/aac608}{{\em Class.
  Quant. Grav.} {\bfseries 35} no.~16, (2018) 163001},
  \href{http://arxiv.org/abs/1801.04268}{{\ttfamily arXiv:1801.04268
  [astro-ph.CO]}}.

\bibitem{Caprini:2019egz}
C.~Caprini {\em et~al.}, ``{Detecting gravitational waves from cosmological
  phase transitions with LISA: an update},''
  \href{http://dx.doi.org/10.1088/1475-7516/2020/03/024}{{\em JCAP} {\bfseries
  2003} (2020) 024},
\href{http://arxiv.org/abs/1910.13125}{{\ttfamily arXiv:1910.13125
  [astro-ph.CO]}}.
%%CITATION = ARXIV:1910.13125;%%.

\bibitem{Athron:2023xlk}
P.~Athron, C.~Bal\'azs, A.~Fowlie, L.~Morris, and L.~Wu, ``{Cosmological phase
  transitions: from perturbative particle physics to gravitational waves},''
  \href{http://arxiv.org/abs/2305.02357}{{\ttfamily arXiv:2305.02357
  [hep-ph]}}.

\bibitem{Randall:2006py}
L.~Randall and G.~Servant, ``{Gravitational waves from warped spacetime},''
  \href{http://dx.doi.org/10.1088/1126-6708/2007/05/054}{{\em JHEP} {\bfseries
  05} (2007) 054}, \href{http://arxiv.org/abs/hep-ph/0607158}{{\ttfamily
  arXiv:hep-ph/0607158}}.

\bibitem{Schwaller:2015tja}
P.~Schwaller, ``{Gravitational Waves from a Dark Phase Transition},''
  \href{http://dx.doi.org/10.1103/PhysRevLett.115.181101}{{\em Phys. Rev.
  Lett.} {\bfseries 115} no.~18, (2015) 181101},
  \href{http://arxiv.org/abs/1504.07263}{{\ttfamily arXiv:1504.07263
  [hep-ph]}}.

\bibitem{Jaeckel:2016jlh}
J.~Jaeckel, V.~V. Khoze, and M.~Spannowsky, ``{Hearing the signal of dark
  sectors with gravitational wave detectors},''
  \href{http://dx.doi.org/10.1103/PhysRevD.94.103519}{{\em Phys. Rev. D}
  {\bfseries 94} no.~10, (2016) 103519},
  \href{http://arxiv.org/abs/1602.03901}{{\ttfamily arXiv:1602.03901
  [hep-ph]}}.

\bibitem{Dev:2016feu}
P.~S.~B. Dev and A.~Mazumdar, ``{Probing the Scale of New Physics by Advanced
  LIGO/VIRGO},'' \href{http://dx.doi.org/10.1103/PhysRevD.93.104001}{{\em Phys.
  Rev. D} {\bfseries 93} no.~10, (2016) 104001},
  \href{http://arxiv.org/abs/1602.04203}{{\ttfamily arXiv:1602.04203
  [hep-ph]}}.

\bibitem{Baldes:2017rcu}
I.~Baldes, ``{Gravitational waves from the asymmetric-dark-matter generating
  phase transition},''
  \href{http://dx.doi.org/10.1088/1475-7516/2017/05/028}{{\em JCAP} {\bfseries
  05} (2017) 028}, \href{http://arxiv.org/abs/1702.02117}{{\ttfamily
  arXiv:1702.02117 [hep-ph]}}.

\bibitem{Tsumura:2017knk}
K.~Tsumura, M.~Yamada, and Y.~Yamaguchi, ``{Gravitational wave from dark sector
  with dark pion},''
  \href{http://dx.doi.org/10.1088/1475-7516/2017/07/044}{{\em JCAP} {\bfseries
  07} (2017) 044}, \href{http://arxiv.org/abs/1704.00219}{{\ttfamily
  arXiv:1704.00219 [hep-ph]}}.

\bibitem{Okada:2018xdh}
N.~Okada and O.~Seto, ``{Probing the seesaw scale with gravitational waves},''
  \href{http://dx.doi.org/10.1103/PhysRevD.98.063532}{{\em Phys. Rev. D}
  {\bfseries 98} no.~6, (2018) 063532},
  \href{http://arxiv.org/abs/1807.00336}{{\ttfamily arXiv:1807.00336
  [hep-ph]}}.

\bibitem{Croon:2018erz}
D.~Croon, V.~Sanz, and G.~White, ``{Model Discrimination in Gravitational Wave
  spectra from Dark Phase Transitions},''
  \href{http://dx.doi.org/10.1007/JHEP08(2018)203}{{\em JHEP} {\bfseries 08}
  (2018) 203}, \href{http://arxiv.org/abs/1806.02332}{{\ttfamily
  arXiv:1806.02332 [hep-ph]}}.

\bibitem{Baldes:2018emh}
I.~Baldes and C.~Garcia-Cely, ``{Strong gravitational radiation from a simple
  dark matter model},'' \href{http://dx.doi.org/10.1007/JHEP05(2019)190}{{\em
  JHEP} {\bfseries 05} (2019) 190},
  \href{http://arxiv.org/abs/1809.01198}{{\ttfamily arXiv:1809.01198
  [hep-ph]}}.

\bibitem{Prokopec:2018tnq}
T.~Prokopec, J.~Rezacek, and B.~\'Swie\.zewska, ``{Gravitational waves from
  conformal symmetry breaking},''
  \href{http://dx.doi.org/10.1088/1475-7516/2019/02/009}{{\em JCAP} {\bfseries
  02} (2019) 009}, \href{http://arxiv.org/abs/1809.11129}{{\ttfamily
  arXiv:1809.11129 [hep-ph]}}.

\bibitem{Bai:2018dxf}
Y.~Bai, A.~J. Long, and S.~Lu, ``{Dark Quark Nuggets},''
  \href{http://dx.doi.org/10.1103/PhysRevD.99.055047}{{\em Phys. Rev. D}
  {\bfseries 99} no.~5, (2019) 055047},
  \href{http://arxiv.org/abs/1810.04360}{{\ttfamily arXiv:1810.04360
  [hep-ph]}}.

\bibitem{Breitbach:2018ddu}
M.~Breitbach, J.~Kopp, E.~Madge, T.~Opferkuch, and P.~Schwaller, ``{Dark, Cold,
  and Noisy: Constraining Secluded Hidden Sectors with Gravitational Waves},''
  \href{http://dx.doi.org/10.1088/1475-7516/2019/07/007}{{\em JCAP} {\bfseries
  07} (2019) 007}, \href{http://arxiv.org/abs/1811.11175}{{\ttfamily
  arXiv:1811.11175 [hep-ph]}}.

\bibitem{Fairbairn:2019xog}
M.~Fairbairn, E.~Hardy, and A.~Wickens, ``{Hearing without seeing:
  gravitational waves from hot and cold hidden sectors},''
  \href{http://dx.doi.org/10.1007/JHEP07(2019)044}{{\em JHEP} {\bfseries 07}
  (2019) 044}, \href{http://arxiv.org/abs/1901.11038}{{\ttfamily
  arXiv:1901.11038 [hep-ph]}}.

\bibitem{Helmboldt:2019pan}
A.~J. Helmboldt, J.~Kubo, and S.~van~der Woude, ``{Observational prospects for
  gravitational waves from hidden or dark chiral phase transitions},''
  \href{http://dx.doi.org/10.1103/PhysRevD.100.055025}{{\em Phys. Rev. D}
  {\bfseries 100} no.~5, (2019) 055025},
  \href{http://arxiv.org/abs/1904.07891}{{\ttfamily arXiv:1904.07891
  [hep-ph]}}.

\bibitem{Ertas:2021xeh}
F.~Ertas, F.~Kahlhoefer, and C.~Tasillo, ``{Turn up the volume: listening to
  phase transitions in hot dark sectors},''
  \href{http://dx.doi.org/10.1088/1475-7516/2022/02/014}{{\em JCAP} {\bfseries
  02} no.~02, (2022) 014}, \href{http://arxiv.org/abs/2109.06208}{{\ttfamily
  arXiv:2109.06208 [astro-ph.CO]}}.

\bibitem{Jinno:2022fom}
R.~Jinno, B.~Shakya, and J.~van~de Vis, ``{Gravitational Waves from Feebly
  Interacting Particles in a First Order Phase Transition},''  (11, 2022) ,
  \href{http://arxiv.org/abs/2211.06405}{{\ttfamily arXiv:2211.06405 [gr-qc]}}.

\bibitem{Baker:2019ndr}
M.~J. Baker, J.~Kopp, and A.~J. Long, ``{Filtered Dark Matter at a First Order
  Phase Transition},''
\href{http://arxiv.org/abs/1912.02830}{{\ttfamily arXiv:1912.02830 [hep-ph]}}.
%%CITATION = ARXIV:1912.02830;%%.

\bibitem{Chway:2019kft}
D.~Chway, T.~H. Jung, and C.~S. Shin, ``{Dark matter filtering-out effect
  during a first-order phase transition},''
\href{http://arxiv.org/abs/1912.04238}{{\ttfamily arXiv:1912.04238 [hep-ph]}}.
%%CITATION = ARXIV:1912.04238;%%.

\bibitem{Asadi:2021yml}
P.~Asadi, E.~D. Kramer, E.~Kuflik, G.~W. Ridgway, T.~R. Slatyer, and
  J.~Smirnov, ``{Accidentally Asymmetric Dark Matter},''
  \href{http://dx.doi.org/10.1103/PhysRevLett.127.211101}{{\em Phys. Rev.
  Lett.} {\bfseries 127} no.~21, (2021) 211101},
  \href{http://arxiv.org/abs/2103.09822}{{\ttfamily arXiv:2103.09822
  [hep-ph]}}.

\bibitem{Asadi:2021pwo}
P.~Asadi, E.~D. Kramer, E.~Kuflik, G.~W. Ridgway, T.~R. Slatyer, and
  J.~Smirnov, ``{Thermal squeezeout of dark matter},''
  \href{http://dx.doi.org/10.1103/PhysRevD.104.095013}{{\em Phys. Rev. D}
  {\bfseries 104} no.~9, (2021) 095013},
  \href{http://arxiv.org/abs/2103.09827}{{\ttfamily arXiv:2103.09827
  [hep-ph]}}.

\bibitem{Asadi:2022vkc}
P.~Asadi, E.~D. Kramer, E.~Kuflik, T.~R. Slatyer, and J.~Smirnov, ``{Glueballs
  in a thermal squeezeout model},''
  \href{http://dx.doi.org/10.1007/JHEP07(2022)006}{{\em JHEP} {\bfseries 07}
  (2022) 006}, \href{http://arxiv.org/abs/2203.15813}{{\ttfamily
  arXiv:2203.15813 [hep-ph]}}.

\bibitem{Gehrman:2023qjn}
T.~C. Gehrman, B.~Shams Es~Haghi, K.~Sinha, and T.~Xu, ``{Recycled dark
  matter},'' \href{http://dx.doi.org/10.1088/1475-7516/2024/03/044}{{\em JCAP}
  {\bfseries 03} (2024) 044}, \href{http://arxiv.org/abs/2310.08526}{{\ttfamily
  arXiv:2310.08526 [hep-ph]}}.

\bibitem{Azatov:2021ifm}
A.~Azatov, M.~Vanvlasselaer, and W.~Yin, ``{Dark Matter production from
  relativistic bubble walls},''
  \href{http://dx.doi.org/10.1007/JHEP03(2021)288}{{\em JHEP} {\bfseries 03}
  (2021) 288}, \href{http://arxiv.org/abs/2101.05721}{{\ttfamily
  arXiv:2101.05721 [hep-ph]}}.

\bibitem{Baldes:2022oev}
I.~Baldes, Y.~Gouttenoire, and F.~Sala, ``{Hot and heavy dark matter from a
  weak scale phase transition},''
  \href{http://dx.doi.org/10.21468/SciPostPhys.14.3.033}{{\em SciPost Phys.}
  {\bfseries 14} no.~3, (2023) 033},
  \href{http://arxiv.org/abs/2207.05096}{{\ttfamily arXiv:2207.05096
  [hep-ph]}}.

\bibitem{Ai:2024ikj}
W.-Y. Ai, M.~Fairbairn, K.~Mimasu, and T.~You, ``{Non-thermal production of
  heavy vector dark matter from relativistic bubble walls},''
  \href{http://arxiv.org/abs/2406.20051}{{\ttfamily arXiv:2406.20051
  [hep-ph]}}.

\bibitem{Hambye:2018qjv}
T.~Hambye, A.~Strumia, and D.~Teresi, ``{Super-cool Dark Matter},''
  \href{http://dx.doi.org/10.1007/JHEP08(2018)188}{{\em JHEP} {\bfseries 08}
  (2018) 188}, \href{http://arxiv.org/abs/1805.01473}{{\ttfamily
  arXiv:1805.01473 [hep-ph]}}.

\bibitem{Baratella:2018pxi}
P.~Baratella, A.~Pomarol, and F.~Rompineve, ``{The Supercooled Universe},''
  \href{http://dx.doi.org/10.1007/JHEP03(2019)100}{{\em JHEP} {\bfseries 03}
  (2019) 100}, \href{http://arxiv.org/abs/1812.06996}{{\ttfamily
  arXiv:1812.06996 [hep-ph]}}.

\bibitem{Baldes:2021aph}
I.~Baldes, Y.~Gouttenoire, F.~Sala, and G.~Servant, ``{Supercool composite Dark
  Matter beyond 100 TeV},''
  \href{http://dx.doi.org/10.1007/JHEP07(2022)084}{{\em JHEP} {\bfseries 07}
  (2022) 084}, \href{http://arxiv.org/abs/2110.13926}{{\ttfamily
  arXiv:2110.13926 [hep-ph]}}.

\bibitem{Falkowski:2012fb}
A.~Falkowski and J.~M. No, ``{Non-thermal Dark Matter Production from the
  Electroweak Phase Transition: Multi-TeV WIMPs and 'Baby-Zillas'},''
  \href{http://dx.doi.org/10.1007/JHEP02(2013)034}{{\em JHEP} {\bfseries 02}
  (2013) 034},
\href{http://arxiv.org/abs/1211.5615}{{\ttfamily arXiv:1211.5615 [hep-ph]}}.
%%CITATION = ARXIV:1211.5615;%%.

\bibitem{Freese:2023fcr}
K.~Freese and M.~W. Winkler, ``{Dark matter and gravitational waves from a dark
  big bang},'' \href{http://dx.doi.org/10.1103/PhysRevD.107.083522}{{\em Phys.
  Rev. D} {\bfseries 107} no.~8, (2023) 083522},
  \href{http://arxiv.org/abs/2302.11579}{{\ttfamily arXiv:2302.11579
  [astro-ph.CO]}}.

\bibitem{Baldes:2023fsp}
I.~Baldes, M.~Dichtl, Y.~Gouttenoire, and F.~Sala, ``{Bubbletrons},''
  \href{http://arxiv.org/abs/2306.15555}{{\ttfamily arXiv:2306.15555
  [hep-ph]}}.

\bibitem{Parker:1969au}
L.~Parker, ``{Quantized fields and particle creation in expanding universes.
  1.},'' \href{http://dx.doi.org/10.1103/PhysRev.183.1057}{{\em Phys. Rev.}
  {\bfseries 183} (1969) 1057--1068}.

\bibitem{Grib:1969ruc}
A.~A. Grib and S.~G. Mamaev, ``{On field theory in the friedman space},'' {\em
  Yad. Fiz.} {\bfseries 10} (1969) 1276--1281.

\bibitem{Zeldovich:1971mw}
Y.~B. Zeldovich and A.~A. Starobinsky, ``{Particle production and vacuum
  polarization in an anisotropic gravitational field},'' {\em Zh. Eksp. Teor.
  Fiz.} {\bfseries 61} (1971) 2161--2175.

\bibitem{Kolb:2023ydq}
E.~W. Kolb and A.~J. Long, ``{Cosmological gravitational particle production
  and its implications for cosmological relics},''
  \href{http://arxiv.org/abs/2312.09042}{{\ttfamily arXiv:2312.09042
  [astro-ph.CO]}}.

\bibitem{Aoki:2022dzd}
S.~Aoki, H.~M. Lee, A.~G. Menkara, and K.~Yamashita, ``{Reheating and dark
  matter freeze-in in the Higgs-R$^{2}$ inflation model},''
  \href{http://dx.doi.org/10.1007/JHEP05(2022)121}{{\em JHEP} {\bfseries 05}
  (2022) 121}, \href{http://arxiv.org/abs/2202.13063}{{\ttfamily
  arXiv:2202.13063 [hep-ph]}}.

\bibitem{Lee:2023dcy}
H.~M. Lee and A.~G. Menkara, ``{Graceful exit from inflation and reheating with
  twin waterfall scalar fields},''
  \href{http://dx.doi.org/10.1103/PhysRevD.107.115019}{{\em Phys. Rev. D}
  {\bfseries 107} no.~11, (2023) 115019},
  \href{http://arxiv.org/abs/2304.08686}{{\ttfamily arXiv:2304.08686
  [hep-ph]}}.

\bibitem{Schwinger:1951nm}
J.~S. Schwinger, ``{On gauge invariance and vacuum polarization},''
  \href{http://dx.doi.org/10.1103/PhysRev.82.664}{{\em Phys. Rev.} {\bfseries
  82} (1951) 664--679}.

\bibitem{Hawking:1974rv}
S.~W. Hawking, ``{Black hole explosions},''
  \href{http://dx.doi.org/10.1038/248030a0}{{\em Nature} {\bfseries 248} (1974)
  30--31}.

\bibitem{Hawking:1975vcx}
S.~W. Hawking, ``{Particle Creation by Black Holes},''
  \href{http://dx.doi.org/10.1007/BF02345020}{{\em Commun. Math. Phys.}
  {\bfseries 43} (1975) 199--220}. [Erratum: Commun.Math.Phys. 46, 206 (1976)].

\bibitem{Watkins:1991zt}
R.~Watkins and L.~M. Widrow, ``{Aspects of reheating in first order
  inflation},''
\href{http://dx.doi.org/10.1016/0550-3213(92)90362-F}{{\em Nucl. Phys.}
  {\bfseries B374} (1992) 446--468}.
%%CITATION = NUPHA,B374,446;%%.

\bibitem{Konstandin:2011ds}
T.~Konstandin and G.~Servant, ``{Natural Cold Baryogenesis from Strongly
  Interacting Electroweak Symmetry Breaking},''
  \href{http://dx.doi.org/10.1088/1475-7516/2011/07/024}{{\em JCAP} {\bfseries
  07} (2011) 024}, \href{http://arxiv.org/abs/1104.4793}{{\ttfamily
  arXiv:1104.4793 [hep-ph]}}.

\bibitem{Mansour:2023fwj}
H.~Mansour and B.~Shakya, ``{On Particle Production from Phase Transition
  Bubbles},'' \href{http://arxiv.org/abs/2308.13070}{{\ttfamily
  arXiv:2308.13070 [hep-ph]}}.

\bibitem{Shakya:2023kjf}
B.~Shakya, ``{Aspects of Particle Production from Bubble Dynamics at a First
  Order Phase Transition},'' \href{http://arxiv.org/abs/2308.16224}{{\ttfamily
  arXiv:2308.16224 [hep-ph]}}.

\bibitem{Inomata:2024rkt}
K.~Inomata, M.~Kamionkowski, K.~Kasai, and B.~Shakya, ``{Gravitational Waves
  from Particles Produced from Bubble Collisions in First-Order Phase
  Transitions},'' \href{http://arxiv.org/abs/2412.17912}{{\ttfamily
  arXiv:2412.17912 [astro-ph.CO]}}.

\bibitem{Carney:2022gse}
D.~Carney {\em et~al.}, ``{Snowmass2021 Cosmic Frontier White Paper: Ultraheavy
  particle dark matter},'' \href{http://arxiv.org/abs/2203.06508}{{\ttfamily
  arXiv:2203.06508 [hep-ph]}}.

\bibitem{McDonald:2001vt}
J.~McDonald, ``{Thermally generated gauge singlet scalars as selfinteracting
  dark matter},'' \href{http://dx.doi.org/10.1103/PhysRevLett.88.091304}{{\em
  Phys. Rev. Lett.} {\bfseries 88} (2002) 091304},
  \href{http://arxiv.org/abs/hep-ph/0106249}{{\ttfamily arXiv:hep-ph/0106249}}.

\bibitem{Hall:2009bx}
L.~J. Hall, K.~Jedamzik, J.~March-Russell, and S.~M. West, ``{Freeze-In
  Production of FIMP Dark Matter},''
  \href{http://dx.doi.org/10.1007/JHEP03(2010)080}{{\em JHEP} {\bfseries 03}
  (2010) 080}, \href{http://arxiv.org/abs/0911.1120}{{\ttfamily arXiv:0911.1120
  [hep-ph]}}.

\bibitem{Bodeker:2009qy}
D.~Bodeker and G.~D. Moore, ``{Can electroweak bubble walls run away?},''
  \href{http://dx.doi.org/10.1088/1475-7516/2009/05/009}{{\em JCAP} {\bfseries
  05} (2009) 009}, \href{http://arxiv.org/abs/0903.4099}{{\ttfamily
  arXiv:0903.4099 [hep-ph]}}.

\bibitem{Dorsch:2018pat}
G.~C. Dorsch, S.~J. Huber, and T.~Konstandin, ``{Bubble wall velocities in the
  Standard Model and beyond},''
  \href{http://dx.doi.org/10.1088/1475-7516/2018/12/034}{{\em JCAP} {\bfseries
  12} (2018) 034}, \href{http://arxiv.org/abs/1809.04907}{{\ttfamily
  arXiv:1809.04907 [hep-ph]}}.

\bibitem{Espinosa:2010hh}
J.~R. Espinosa, T.~Konstandin, J.~M. No, and G.~Servant, ``{Energy Budget of
  Cosmological First-order Phase Transitions},''
  \href{http://dx.doi.org/10.1088/1475-7516/2010/06/028}{{\em JCAP} {\bfseries
  06} (2010) 028}, \href{http://arxiv.org/abs/1004.4187}{{\ttfamily
  arXiv:1004.4187 [hep-ph]}}.

\bibitem{Ai:2024shx}
W.-Y. Ai, X.~Nagels, and M.~Vanvlasselaer, ``{Criterion for ultra-fast bubble
  walls: the impact of hydrodynamic obstruction},''
  \href{http://dx.doi.org/10.1088/1475-7516/2024/03/037}{{\em JCAP} {\bfseries
  03} (2024) 037}, \href{http://arxiv.org/abs/2401.05911}{{\ttfamily
  arXiv:2401.05911 [hep-ph]}}.

\bibitem{Bodeker:2017cim}
D.~Bodeker and G.~D. Moore, ``{Electroweak Bubble Wall Speed Limit},''
  \href{http://dx.doi.org/10.1088/1475-7516/2017/05/025}{{\em JCAP} {\bfseries
  05} (2017) 025}, \href{http://arxiv.org/abs/1703.08215}{{\ttfamily
  arXiv:1703.08215 [hep-ph]}}.

\bibitem{Hoche:2020ysm}
S.~H\"oche, J.~Kozaczuk, A.~J. Long, J.~Turner, and Y.~Wang, ``{Towards an
  all-orders calculation of the electroweak bubble wall velocity},''
  \href{http://dx.doi.org/10.1088/1475-7516/2021/03/009}{{\em JCAP} {\bfseries
  03} (2021) 009}, \href{http://arxiv.org/abs/2007.10343}{{\ttfamily
  arXiv:2007.10343 [hep-ph]}}.

\bibitem{Gouttenoire:2021kjv}
Y.~Gouttenoire, R.~Jinno, and F.~Sala, ``{Friction pressure on relativistic
  bubble walls},''  (12, 2021) ,
  \href{http://arxiv.org/abs/2112.07686}{{\ttfamily arXiv:2112.07686
  [hep-ph]}}.

\bibitem{Ellis:2020nnr}
J.~Ellis, M.~Lewicki, and V.~Vaskonen, ``{Updated predictions for gravitational
  waves produced in a strongly supercooled phase transition},''
  \href{http://arxiv.org/abs/2007.15586}{{\ttfamily arXiv:2007.15586
  [astro-ph.CO]}}.

\bibitem{Konstandin:2011dr}
T.~Konstandin and G.~Servant, ``{Cosmological Consequences of Nearly Conformal
  Dynamics at the TeV scale},''
  \href{http://dx.doi.org/10.1088/1475-7516/2011/12/009}{{\em JCAP} {\bfseries
  12} (2011) 009}, \href{http://arxiv.org/abs/1104.4791}{{\ttfamily
  arXiv:1104.4791 [hep-ph]}}.

\bibitem{vonHarling:2017yew}
B.~von Harling and G.~Servant, ``{QCD-induced Electroweak Phase Transition},''
  \href{http://dx.doi.org/10.1007/JHEP01(2018)159}{{\em JHEP} {\bfseries 01}
  (2018) 159}, \href{http://arxiv.org/abs/1711.11554}{{\ttfamily
  arXiv:1711.11554 [hep-ph]}}.

\bibitem{Ellis:2018mja}
J.~Ellis, M.~Lewicki, and J.~M. No, ``{On the Maximal Strength of a First-Order
  Electroweak Phase Transition and its Gravitational Wave Signal},''
  \href{http://dx.doi.org/10.1088/1475-7516/2019/04/003}{{\em JCAP} {\bfseries
  04} (2019) 003}, \href{http://arxiv.org/abs/1809.08242}{{\ttfamily
  arXiv:1809.08242 [hep-ph]}}.

\bibitem{Bruggisser:2018mrt}
S.~Bruggisser, B.~Von~Harling, O.~Matsedonskyi, and G.~Servant, ``{Electroweak
  Phase Transition and Baryogenesis in Composite Higgs Models},''
  \href{http://dx.doi.org/10.1007/JHEP12(2018)099}{{\em JHEP} {\bfseries 12}
  (2018) 099}, \href{http://arxiv.org/abs/1804.07314}{{\ttfamily
  arXiv:1804.07314 [hep-ph]}}.

\bibitem{DelleRose:2019pgi}
L.~Delle~Rose, G.~Panico, M.~Redi, and A.~Tesi, ``{Gravitational Waves from
  Supercool Axions},'' \href{http://dx.doi.org/10.1007/JHEP04(2020)025}{{\em
  JHEP} {\bfseries 04} (2020) 025},
  \href{http://arxiv.org/abs/1912.06139}{{\ttfamily arXiv:1912.06139
  [hep-ph]}}.

\bibitem{VonHarling:2019rgb}
B.~Von~Harling, A.~Pomarol, O.~Pujol\`as, and F.~Rompineve, ``{Peccei-Quinn
  Phase Transition at LIGO},''
  \href{http://dx.doi.org/10.1007/JHEP04(2020)195}{{\em JHEP} {\bfseries 04}
  (2020) 195}, \href{http://arxiv.org/abs/1912.07587}{{\ttfamily
  arXiv:1912.07587 [hep-ph]}}.

\bibitem{Fujikura:2019oyi}
K.~Fujikura, Y.~Nakai, and M.~Yamada, ``{A more attractive scheme for radion
  stabilization and supercooled phase transition},''
  \href{http://dx.doi.org/10.1007/JHEP02(2020)111}{{\em JHEP} {\bfseries 02}
  (2020) 111}, \href{http://arxiv.org/abs/1910.07546}{{\ttfamily
  arXiv:1910.07546 [hep-ph]}}.

\bibitem{Ellis:2019oqb}
J.~Ellis, M.~Lewicki, J.~M. No, and V.~Vaskonen, ``{Gravitational wave energy
  budget in strongly supercooled phase transitions},''
  \href{http://dx.doi.org/10.1088/1475-7516/2019/06/024}{{\em JCAP} {\bfseries
  06} (2019) 024}, \href{http://arxiv.org/abs/1903.09642}{{\ttfamily
  arXiv:1903.09642 [hep-ph]}}.

\bibitem{Brdar:2019qut}
V.~Brdar, A.~J. Helmboldt, and M.~Lindner, ``{Strong Supercooling as a
  Consequence of Renormalization Group Consistency},''
  \href{http://dx.doi.org/10.1007/JHEP12(2019)158}{{\em JHEP} {\bfseries 12}
  (2019) 158}, \href{http://arxiv.org/abs/1910.13460}{{\ttfamily
  arXiv:1910.13460 [hep-ph]}}.

\bibitem{Baldes:2020kam}
I.~Baldes, Y.~Gouttenoire, and F.~Sala, ``{String Fragmentation in Supercooled
  Confinement and Implications for Dark Matter},''
  \href{http://dx.doi.org/10.1007/JHEP04(2021)278}{{\em JHEP} {\bfseries 04}
  (2021) 278}, \href{http://arxiv.org/abs/2007.08440}{{\ttfamily
  arXiv:2007.08440 [hep-ph]}}.

\bibitem{Bruggisser:2022rdm}
S.~Bruggisser, B.~von Harling, O.~Matsedonskyi, and G.~Servant, ``{Status of
  electroweak baryogenesis in minimal composite Higgs},''
  \href{http://dx.doi.org/10.1007/JHEP08(2023)012}{{\em JHEP} {\bfseries 08}
  (2023) 012}, \href{http://arxiv.org/abs/2212.11953}{{\ttfamily
  arXiv:2212.11953 [hep-ph]}}.

\bibitem{Cutting:2020nla}
D.~Cutting, E.~G. Escartin, M.~Hindmarsh, and D.~J. Weir, ``{Gravitational
  waves from vacuum first order phase transitions II: from thin to thick
  walls},'' \href{http://dx.doi.org/10.1103/PhysRevD.103.023531}{{\em Phys.
  Rev. D} {\bfseries 103} no.~2, (2021) 023531},
  \href{http://arxiv.org/abs/2005.13537}{{\ttfamily arXiv:2005.13537
  [astro-ph.CO]}}.

\bibitem{Hawking:1982ga}
S.~W. Hawking, I.~G. Moss, and J.~M. Stewart, ``{Bubble Collisions in the Very
  Early Universe},'' \href{http://dx.doi.org/10.1103/PhysRevD.26.2681}{{\em
  Phys. Rev. D} {\bfseries 26} (1982) 2681}.

\bibitem{Cuomo:2019siu}
G.~Cuomo, L.~Vecchi, and A.~Wulzer, ``{Goldstone Equivalence and High Energy
  Electroweak Physics},''
  \href{http://dx.doi.org/10.21468/SciPostPhys.8.5.078}{{\em SciPost Phys.}
  {\bfseries 8} no.~5, (2020) 078},
  \href{http://arxiv.org/abs/1911.12366}{{\ttfamily arXiv:1911.12366
  [hep-ph]}}.

\bibitem{Wulzer:2013mza}
A.~Wulzer, ``{An Equivalent Gauge and the Equivalence Theorem},''
  \href{http://dx.doi.org/10.1016/j.nuclphysb.2014.05.021}{{\em Nucl. Phys. B}
  {\bfseries 885} (2014) 97--126},
  \href{http://arxiv.org/abs/1309.6055}{{\ttfamily arXiv:1309.6055 [hep-ph]}}.

\bibitem{Papavassiliou:1997fn}
J.~Papavassiliou and A.~Pilaftsis, ``{Effective charge of the Higgs boson},''
  \href{http://dx.doi.org/10.1103/PhysRevLett.80.2785}{{\em Phys. Rev. Lett.}
  {\bfseries 80} (1998) 2785--2788},
  \href{http://arxiv.org/abs/hep-ph/9710380}{{\ttfamily arXiv:hep-ph/9710380}}.

\bibitem{Papavassiliou:1997pb}
J.~Papavassiliou and A.~Pilaftsis, ``{Gauge and renormalization group invariant
  formulation of the Higgs boson resonance},''
  \href{http://dx.doi.org/10.1103/PhysRevD.58.053002}{{\em Phys. Rev. D}
  {\bfseries 58} (1998) 053002},
  \href{http://arxiv.org/abs/hep-ph/9710426}{{\ttfamily arXiv:hep-ph/9710426}}.

\bibitem{Duch:2018ucs}
M.~Duch, B.~Grzadkowski, and A.~Pilaftsis, ``{Gauge-Independent Approach to
  Resonant Dark Matter Annihilation},''
  \href{http://dx.doi.org/10.1007/JHEP02(2019)141}{{\em JHEP} {\bfseries 02}
  (2019) 141}, \href{http://arxiv.org/abs/1812.11944}{{\ttfamily
  arXiv:1812.11944 [hep-ph]}}.

\bibitem{Giudice:2000ex}
G.~F. Giudice, E.~W. Kolb, and A.~Riotto, ``{Largest temperature of the
  radiation era and its cosmological implications},''
  \href{http://dx.doi.org/10.1103/PhysRevD.64.023508}{{\em Phys. Rev. D}
  {\bfseries 64} (2001) 023508},
  \href{http://arxiv.org/abs/hep-ph/0005123}{{\ttfamily arXiv:hep-ph/0005123}}.

\bibitem{Chung:1998rq}
D.~J.~H. Chung, E.~W. Kolb, and A.~Riotto, ``{Production of massive particles
  during reheating},'' \href{http://dx.doi.org/10.1103/PhysRevD.60.063504}{{\em
  Phys. Rev. D} {\bfseries 60} (1999) 063504},
  \href{http://arxiv.org/abs/hep-ph/9809453}{{\ttfamily arXiv:hep-ph/9809453}}.

\bibitem{Frangipane:2021rtf}
E.~Frangipane, S.~Gori, and B.~Shakya, ``{Dark matter freeze-in with a heavy
  mediator: beyond the EFT approach},''
  \href{http://dx.doi.org/10.1007/JHEP09(2022)083}{{\em JHEP} {\bfseries 09}
  (2022) 083}, \href{http://arxiv.org/abs/2110.10711}{{\ttfamily
  arXiv:2110.10711 [hep-ph]}}.

\bibitem{Elahi:2014fsa}
F.~Elahi, C.~Kolda, and J.~Unwin, ``{UltraViolet Freeze-in},''
  \href{http://dx.doi.org/10.1007/JHEP03(2015)048}{{\em JHEP} {\bfseries 03}
  (2015) 048}, \href{http://arxiv.org/abs/1410.6157}{{\ttfamily arXiv:1410.6157
  [hep-ph]}}.

\bibitem{Roland:2016gli}
S.~B. Roland and B.~Shakya, ``{Cosmological Imprints of Frozen-In Light Sterile
  Neutrinos},'' \href{http://dx.doi.org/10.1088/1475-7516/2017/05/027}{{\em
  JCAP} {\bfseries 05} (2017) 027},
  \href{http://arxiv.org/abs/1609.06739}{{\ttfamily arXiv:1609.06739
  [hep-ph]}}.

\bibitem{Azatov:2023xem}
A.~Azatov, G.~Barni, R.~Petrossian-Byrne, and M.~Vanvlasselaer, ``{Quantisation
  Across Bubble Walls and Friction},''
  \href{http://arxiv.org/abs/2310.06972}{{\ttfamily arXiv:2310.06972
  [hep-ph]}}.

\bibitem{Azatov:2020ufh}
A.~Azatov and M.~Vanvlasselaer, ``{Bubble wall velocity: heavy physics
  effects},'' \href{http://dx.doi.org/10.1088/1475-7516/2021/01/058}{{\em JCAP}
  {\bfseries 01} (2021) 058}, \href{http://arxiv.org/abs/2010.02590}{{\ttfamily
  arXiv:2010.02590 [hep-ph]}}.

\bibitem{Azatov:2022tii}
A.~Azatov, G.~Barni, S.~Chakraborty, M.~Vanvlasselaer, and W.~Yin,
  ``{Ultra-relativistic bubbles from the simplest Higgs portal and their
  cosmological consequences},''
  \href{http://dx.doi.org/10.1007/JHEP10(2022)017}{{\em JHEP} {\bfseries 10}
  (2022) 017}, \href{http://arxiv.org/abs/2207.02230}{{\ttfamily
  arXiv:2207.02230 [hep-ph]}}.

\bibitem{Baldes:2024wuz}
I.~Baldes, M.~Dichtl, Y.~Gouttenoire, and F.~Sala, ``{Particle shells from
  relativistic bubble walls},''
  \href{http://dx.doi.org/10.1007/JHEP07(2024)231}{{\em JHEP} {\bfseries 07}
  (2024) 231}, \href{http://arxiv.org/abs/2403.05615}{{\ttfamily
  arXiv:2403.05615 [hep-ph]}}.

\bibitem{Shakya:2016oxf}
B.~Shakya and J.~D. Wells, ``{Sterile Neutrino Dark Matter with
  Supersymmetry},'' \href{http://dx.doi.org/10.1103/PhysRevD.96.031702}{{\em
  Phys. Rev. D} {\bfseries 96} no.~3, (2017) 031702},
  \href{http://arxiv.org/abs/1611.01517}{{\ttfamily arXiv:1611.01517
  [hep-ph]}}.

\bibitem{Windchime:2022whs}
{\bfseries Windchime} Collaboration, A.~Attanasio {\em et~al.}, ``{Snowmass
  2021 White Paper: The Windchime Project},'' in {\em {Snowmass 2021}}.
\newblock 3, 2022.
\newblock \href{http://arxiv.org/abs/2203.07242}{{\ttfamily arXiv:2203.07242
  [hep-ex]}}.

\bibitem{Caprini:2007xq}
C.~Caprini, R.~Durrer, and G.~Servant, ``{Gravitational wave generation from
  bubble collisions in first-order phase transitions: An analytic approach},''
  \href{http://dx.doi.org/10.1103/PhysRevD.77.124015}{{\em Phys. Rev. D}
  {\bfseries 77} (2008) 124015},
  \href{http://arxiv.org/abs/0711.2593}{{\ttfamily arXiv:0711.2593
  [astro-ph]}}.

\bibitem{Huber:2008hg}
S.~J. Huber and T.~Konstandin, ``{Gravitational Wave Production by Collisions:
  More Bubbles},'' \href{http://dx.doi.org/10.1088/1475-7516/2008/09/022}{{\em
  JCAP} {\bfseries 09} (2008) 022},
  \href{http://arxiv.org/abs/0806.1828}{{\ttfamily arXiv:0806.1828 [hep-ph]}}.

\bibitem{Jinno:2016vai}
R.~Jinno and M.~Takimoto, ``{Gravitational waves from bubble collisions: An
  analytic derivation},''
  \href{http://dx.doi.org/10.1103/PhysRevD.95.024009}{{\em Phys. Rev. D}
  {\bfseries 95} no.~2, (2017) 024009},
  \href{http://arxiv.org/abs/1605.01403}{{\ttfamily arXiv:1605.01403
  [astro-ph.CO]}}.

\bibitem{Jinno:2017fby}
R.~Jinno and M.~Takimoto, ``{Gravitational waves from bubble dynamics: Beyond
  the Envelope},'' \href{http://dx.doi.org/10.1088/1475-7516/2019/01/060}{{\em
  JCAP} {\bfseries 01} (2019) 060},
  \href{http://arxiv.org/abs/1707.03111}{{\ttfamily arXiv:1707.03111
  [hep-ph]}}.

\bibitem{Konstandin:2017sat}
T.~Konstandin, ``{Gravitational radiation from a bulk flow model},''
  \href{http://dx.doi.org/10.1088/1475-7516/2018/03/047}{{\em JCAP} {\bfseries
  03} (2018) 047}, \href{http://arxiv.org/abs/1712.06869}{{\ttfamily
  arXiv:1712.06869 [astro-ph.CO]}}.

\bibitem{Cutting:2018tjt}
D.~Cutting, M.~Hindmarsh, and D.~J. Weir, ``{Gravitational waves from vacuum
  first-order phase transitions: from the envelope to the lattice},''
  \href{http://dx.doi.org/10.1103/PhysRevD.97.123513}{{\em Phys. Rev. D}
  {\bfseries 97} no.~12, (2018) 123513},
  \href{http://arxiv.org/abs/1802.05712}{{\ttfamily arXiv:1802.05712
  [astro-ph.CO]}}.

\bibitem{Lewicki:2020azd}
M.~Lewicki and V.~Vaskonen, ``{Gravitational waves from colliding vacuum
  bubbles in gauge theories},''
  \href{http://dx.doi.org/10.1140/epjc/s10052-021-09232-3}{{\em Eur. Phys. J.
  C} {\bfseries 81} no.~5, (2021) 437},
  \href{http://arxiv.org/abs/2012.07826}{{\ttfamily arXiv:2012.07826
  [astro-ph.CO]}}. [Erratum: Eur.Phys.J.C 81, 1077 (2021)].

\bibitem{Hindmarsh:2013xza}
M.~Hindmarsh, S.~J. Huber, K.~Rummukainen, and D.~J. Weir, ``{Gravitational
  waves from the sound of a first order phase transition},''
  \href{http://dx.doi.org/10.1103/PhysRevLett.112.041301}{{\em Phys. Rev.
  Lett.} {\bfseries 112} (2014) 041301},
  \href{http://arxiv.org/abs/1304.2433}{{\ttfamily arXiv:1304.2433 [hep-ph]}}.

\bibitem{Hindmarsh:2015qta}
M.~Hindmarsh, S.~J. Huber, K.~Rummukainen, and D.~J. Weir, ``{Numerical
  simulations of acoustically generated gravitational waves at a first order
  phase transition},'' \href{http://dx.doi.org/10.1103/PhysRevD.92.123009}{{\em
  Phys. Rev. D} {\bfseries 92} no.~12, (2015) 123009},
  \href{http://arxiv.org/abs/1504.03291}{{\ttfamily arXiv:1504.03291
  [astro-ph.CO]}}.

\bibitem{Hindmarsh:2017gnf}
M.~Hindmarsh, S.~J. Huber, K.~Rummukainen, and D.~J. Weir, ``{Shape of the
  acoustic gravitational wave power spectrum from a first order phase
  transition},'' \href{http://dx.doi.org/10.1103/PhysRevD.96.103520}{{\em Phys.
  Rev. D} {\bfseries 96} no.~10, (2017) 103520},
  \href{http://arxiv.org/abs/1704.05871}{{\ttfamily arXiv:1704.05871
  [astro-ph.CO]}}. [Erratum: Phys.Rev.D 101, 089902 (2020)].

\bibitem{Cutting:2019zws}
D.~Cutting, M.~Hindmarsh, and D.~J. Weir, ``{Vorticity, kinetic energy, and
  suppressed gravitational wave production in strong first order phase
  transitions},'' \href{http://dx.doi.org/10.1103/PhysRevLett.125.021302}{{\em
  Phys. Rev. Lett.} {\bfseries 125} no.~2, (2020) 021302},
  \href{http://arxiv.org/abs/1906.00480}{{\ttfamily arXiv:1906.00480
  [hep-ph]}}.

\bibitem{Hindmarsh:2016lnk}
M.~Hindmarsh, ``{Sound shell model for acoustic gravitational wave production
  at a first-order phase transition in the early Universe},''
  \href{http://dx.doi.org/10.1103/PhysRevLett.120.071301}{{\em Phys. Rev.
  Lett.} {\bfseries 120} no.~7, (2018) 071301},
  \href{http://arxiv.org/abs/1608.04735}{{\ttfamily arXiv:1608.04735
  [astro-ph.CO]}}.

\bibitem{Hindmarsh:2019phv}
M.~Hindmarsh and M.~Hijazi, ``{Gravitational waves from first order
  cosmological phase transitions in the Sound Shell Model},''
  \href{http://dx.doi.org/10.1088/1475-7516/2019/12/062}{{\em JCAP} {\bfseries
  12} (2019) 062}, \href{http://arxiv.org/abs/1909.10040}{{\ttfamily
  arXiv:1909.10040 [astro-ph.CO]}}.

\bibitem{Caprini:2009yp}
C.~Caprini, R.~Durrer, and G.~Servant, ``{The stochastic gravitational wave
  background from turbulence and magnetic fields generated by a first-order
  phase transition},''
  \href{http://dx.doi.org/10.1088/1475-7516/2009/12/024}{{\em JCAP} {\bfseries
  12} (2009) 024}, \href{http://arxiv.org/abs/0909.0622}{{\ttfamily
  arXiv:0909.0622 [astro-ph.CO]}}.

\bibitem{Brandenburg:2017neh}
A.~Brandenburg, T.~Kahniashvili, S.~Mandal, A.~Roper~Pol, A.~G. Tevzadze, and
  T.~Vachaspati, ``{Evolution of hydromagnetic turbulence from the electroweak
  phase transition},'' \href{http://dx.doi.org/10.1103/PhysRevD.96.123528}{{\em
  Phys. Rev. D} {\bfseries 96} no.~12, (2017) 123528},
  \href{http://arxiv.org/abs/1711.03804}{{\ttfamily arXiv:1711.03804
  [astro-ph.CO]}}.

\bibitem{RoperPol:2019wvy}
A.~Roper~Pol, S.~Mandal, A.~Brandenburg, T.~Kahniashvili, and A.~Kosowsky,
  ``{Numerical simulations of gravitational waves from early-universe
  turbulence},'' \href{http://dx.doi.org/10.1103/PhysRevD.102.083512}{{\em
  Phys. Rev. D} {\bfseries 102} no.~8, (2020) 083512},
  \href{http://arxiv.org/abs/1903.08585}{{\ttfamily arXiv:1903.08585
  [astro-ph.CO]}}.

\bibitem{Dahl:2021wyk}
J.~Dahl, M.~Hindmarsh, K.~Rummukainen, and D.~Weir, ``{Decay of acoustic
  turbulence in two dimensions and implications for cosmological gravitational
  waves},''  (12, 2021) , \href{http://arxiv.org/abs/2112.12013}{{\ttfamily
  arXiv:2112.12013 [gr-qc]}}.

\bibitem{Auclair:2022jod}
P.~Auclair, C.~Caprini, D.~Cutting, M.~Hindmarsh, K.~Rummukainen, D.~A. Steer,
  and D.~J. Weir, ``{Generation of gravitational waves from freely decaying
  turbulence},'' \href{http://dx.doi.org/10.1088/1475-7516/2022/09/029}{{\em
  JCAP} {\bfseries 09} (2022) 029},
  \href{http://arxiv.org/abs/2205.02588}{{\ttfamily arXiv:2205.02588
  [astro-ph.CO]}}.

\bibitem{Bringmann:2023opz}
T.~Bringmann, P.~F. Depta, T.~Konstandin, K.~Schmidt-Hoberg, and C.~Tasillo,
  ``{Does NANOGrav observe a dark sector phase transition?},''
  \href{http://dx.doi.org/10.1088/1475-7516/2023/11/053}{{\em JCAP} {\bfseries
  11} (2023) 053}, \href{http://arxiv.org/abs/2306.09411}{{\ttfamily
  arXiv:2306.09411 [astro-ph.CO]}}.

\bibitem{Madge:2023dxc}
E.~Madge, E.~Morgante, C.~Puchades-Ib\'a\~nez, N.~Ramberg, W.~Ratzinger,
  S.~Schenk, and P.~Schwaller, ``{Primordial gravitational waves in the
  nano-Hertz regime and PTA data \textemdash{} towards solving the GW inverse
  problem},'' \href{http://dx.doi.org/10.1007/JHEP10(2023)171}{{\em JHEP}
  {\bfseries 10} (2023) 171}, \href{http://arxiv.org/abs/2306.14856}{{\ttfamily
  arXiv:2306.14856 [hep-ph]}}.

\bibitem{LISA:2017pwj}
{\bfseries LISA} Collaboration, P.~Amaro-Seoane {\em et~al.}, ``{Laser
  Interferometer Space Antenna},''
  \href{http://arxiv.org/abs/1702.00786}{{\ttfamily arXiv:1702.00786
  [astro-ph.IM]}}.

\bibitem{Harry:2006fi}
G.~M. Harry, P.~Fritschel, D.~A. Shaddock, W.~Folkner, and E.~S. Phinney,
  ``{Laser interferometry for the big bang observer},''
  \href{http://dx.doi.org/10.1088/0264-9381/23/15/008}{{\em Class. Quant.
  Grav.} {\bfseries 23} (2006) 4887--4894}. [Erratum: Class.Quant.Grav. 23,
  7361 (2006)].

\bibitem{Kawamura:2006up}
S.~Kawamura {\em et~al.}, ``{The Japanese space gravitational wave antenna
  DECIGO},'' \href{http://dx.doi.org/10.1088/0264-9381/23/8/S17}{{\em Class.
  Quant. Grav.} {\bfseries 23} (2006) S125--S132}.

\bibitem{Punturo:2010zz}
M.~Punturo {\em et~al.}, ``{The Einstein Telescope: A third-generation
  gravitational wave observatory},''
  \href{http://dx.doi.org/10.1088/0264-9381/27/19/194002}{{\em Class. Quant.
  Grav.} {\bfseries 27} (2010) 194002}.

\bibitem{Reitze:2019iox}
D.~Reitze {\em et~al.}, ``{Cosmic Explorer: The U.S. Contribution to
  Gravitational-Wave Astronomy beyond LIGO},'' {\em Bull. Am. Astron. Soc.}
  {\bfseries 51} no.~7, (2019) 035,
  \href{http://arxiv.org/abs/1907.04833}{{\ttfamily arXiv:1907.04833
  [astro-ph.IM]}}.

\bibitem{Cataldi:2024pgt}
M.~Cataldi and B.~Shakya, ``{Leptogenesis via bubble collisions},''
  \href{http://dx.doi.org/10.1088/1475-7516/2024/11/047}{{\em JCAP} {\bfseries
  11} (2024) 047}, \href{http://arxiv.org/abs/2407.16747}{{\ttfamily
  arXiv:2407.16747 [hep-ph]}}.

\end{thebibliography}\endgroup

\end{document}